\documentclass[authoryear]{elsarticle}
\usepackage[english]{babel}
\usepackage[utf8]{inputenc}
\usepackage{amsmath}
\usepackage{amsthm}
\usepackage{amsfonts,amssymb,amsthm,amsmath}
\usepackage{graphicx}
\usepackage{xcolor}
\usepackage[colorinlistoftodos]{todonotes}
\usepackage{stackengine}
\usepackage{scalerel}
\usepackage{setspace}
\usepackage{graphicx}
\usepackage{subfig}
\newtheorem{proposition}{Proposition}

\newtheorem{definition}{Definition}
\newtheorem{remark}{Remark}
\newtheorem{theorem}{Theorem}

\newtheorem{corollary}{Corollary}
\usepackage{float}
\usepackage{lscape}
\usepackage{setspace}
\usepackage{slashbox}
\usepackage{float}

\newcommand{\phinegrita}{\mbox{\boldmath$\phi$}}

\newcommand{\bmap}{\emph{BMAP}}

\newcommand{\be}{\begin{eqnarray}}
	\newcommand{\ee}{\end{eqnarray}}
\newcommand{\beq}{\begin{equation}}
	\newcommand{\eeq}{\end{equation}}
\newcommand{\ea}{\end{eqnarray*}}
\newcommand{\ba}{\begin{eqnarray}}
\newcommand{\edi}{\end{displaymath}}
\newcommand{\bdi}{\begin{displaymath}}

\newcommand{\map}{\emph{MAP}}
\newcommand{\mmpp}{\emph{MMPP}}
\newcommand{\bmmpp}{\emph{BMMPP}}

\usepackage{colortbl}

\definecolor{maroon}{cmyk}{0,0.87,0.68,0.32}

\usepackage[detect-all]{siunitx}

\usepackage{multirow}
\title{A bivariate two-state Markov modulated Poisson process for failure modelling}
\author[uc3m]{Yoel G. Yera\corref{cor1}}
\ead{yoelgustavo.yera@uc3m.es}
\author[uc3m,UC3M-BS]{Rosa E. Lillo}
\ead{rosaelvira.lillo@uc3m.es}
\author[dtu]{Bo F. Nielsen}
\ead{bfni@dtu.dk}
\author[uca]{Pepa Ramírez-Cobo}
\ead{pepa.ramirez@uca.es}
\author[cnr]{Fabrizio Ruggeri}
\ead{fabrizio@mi.imati.cnr.it}

\cortext[cor1]{Corresponding author}

\address[uc3m]{Department of Statistics, Universidad Carlos III de Madrid, Spain}
\address[UC3M-BS]{uc3m-Santander Big Data Institute, Madrid, Spain}
\address[dtu]{Department of Applied Mathematics and Computer Science. Technical University of Denmark, Kongens Lyngby, Denmark}
\address[uca]{Department of Statistics and Operations Research, Universidad de C\'adiz, Spain}
\address[cnr]{IMATI-CNR	Institute of Applied Mathematics and Information Technologies, Milano, Italy}

\begin{document}

	\begin{abstract}
		Motivated by a real failure dataset in a two-dimensional context, this paper presents an extension of the Markov modulated Poisson process ($\mmpp$) to two dimensions. The one-dimensional $\mmpp$  has been proposed for the modeling of dependent and non-exponential inter-failure times (in contexts as queuing, risk or reliability, among others). The novel two-dimensional $\mmpp$ allows for dependence between the two sequences of inter-failure times, while at the same time preserves the $\mmpp$ properties, marginally. The generalization is based on the Marshall-Olkin exponential distribution. Inference is undertaken for the new model through a method combining a matching moments approach with an \textcolor{black}{Approximate Bayesian Computation (ABC)} algorithm. The performance of the method is shown on simulated and real datasets representing times and distances covered between consecutive failures in a public transport company. For the real dataset, some quantities of importance associated with the reliability of the system are estimated as the probabilities and expected number of failures at different times and distances covered by trains until the occurrence of a failure.
		
	\end{abstract}
	
	
	\maketitle
	\section*{List of acronyms}
	\begin{itemize}
	    \item \textbf{ABC}: Approximate Bayesian computation algorithm
	    \item \textbf{$\bmap$}: Batch Markovian arrival process
	    \item \textbf{$\bmmpp$}: Batch Markov modulated Poisson process
	    \item \textbf{$\bmmpp_2$}: Two-state Batch Markov modulated Poisson process
	    \item \textbf{BVE}: Bivariate exponential distribution
	    \item \textbf{$\map$}: Markovian arrival process
	    \item \textbf{$\map_2$}: Two-state Markovian arrival process
	    \item \textbf{$\mmpp$}: Markov modulated Poisson process
	    \item \textbf{$\mmpp_2$}: Two-state Markov modulated Poisson process
	    \item \textbf{MPH}: Multivariate Phase-Type distribution
	    \item \textbf{SPP}: Switched Poisson Process
	\end{itemize}

	\section{Introduction}\label{sec:bivintro}
	
	Transportation means are essential in every day life and, therefore, it is crucial not only to detect failures but also to analyze their nature and the way to prevent them. Reliability studies associated with aviation, vehicles, rail defects and ships failures have been considered in a number of articles, see for example \cite{zhang2002,zhang2005,karim2008,fang2005,ivanov2009,Deco2012,kim2018integrated,wang2020sensitivity} or \cite{ressnew}, just to cite a few. In particular, with the rapid development of the railway industry and high-speed lines, safety and reliability of train systems have attracted increasing attention.
	As a consequence, a number of studies dealing with trains system failures have been carried out, see for example \cite{navas17,kim2018integrated,chang2020hybrid}, chapters 5, 7 and 10 from \cite{book1}, or more recently, chapter 6 from \cite{book2}. The study by \cite{nelson2000commuter} revealed the distribution of the percentages of delays due to commuter rail equipment failures. In particular, the $20\%$ of delays are commonly caused by prime-mover problems, the $13\%$ are caused by braking system problem and the $7\%$ of delays are caused by problems related to coach components (as doors) in control cabs. Taking as starting point a real two-dimensional dataset representing doors' failures times and covered distances in a lot of underground trains, this paper presents a novel stochastic model that captures the main statistical properties observed from data and is able to provide insight about the system reliability.

	Two-dimensional traces (or two-scale data) are not rare in the field of reliability. For example, in the context of warranty data (automobile, automobile tyres), failures depend on both the age and amount of accumulated usage, see  \cite{eliashberg1997,yang2001bivariate,981654,huang2013two,huang2015cost}. Other examples different from transportation, where two-dimensional models are important, include  factory equipment, power generation machines or aircraft, see \cite{935013}. From a stochastic modeling viewpoint, problems where longevity is measured in terms of two quantities (age and usage) need to be addressed by bivariate models able to capture the existing dependence between the two measures, see \cite{hunter1974renewal,singpurwalla1998failure,pievatolo2003,ruggeri06,pievatolo2010, d2011age,KUMARGUPTA201764}. 
	
	
	In this article we present a novel bivariate stochastic model that is able to provide a good fit for the two-dimensional dataset considered by \cite{pievatolo2003,pievatolo2010}. The dataset consists of  records of the failures in trains doors for a period of 8 years. The final goal in those papers was to detect, before the warranty expired, if the trains were reliable as stated in the contract with the manufacturer. Here reliability is measured in terms of the number of failures for a specified time period or a distance covered. Therefore, when a failure took place, both the reading of the odometer (which quantifies the number of kilometres covered) and the date of the failure were recorded. As described in detail in Section \ref{sec:studytrains}, both components of the considered dataset (inter-failure time and distance) present a high linear dependence. Also, neither the inter-failure times nor the distances covered can be assumed to be generated by an exponential distribution. Finally, both sequences of  inter-failure times and distances present a non-negligible autocorrelation structure. 
	
	
	In analogous circumstances (non-expontiality, non-negligible autocorrelation function) but in univariate case, some authors have suggested the use of the Markovian arrival process ($\map$), see for example \cite{Buchholz2003,Klemm,Ram08,RamirezCoboyCarrizosa2012,Liu15,RodInf,Rod161,BayesPepa}. Within the class of $\map$s, it is worth mentioning the Markov modulated Poisson process (noted $\mmpp$), which constitutes a versatile and identifiable subclass, see \cite{Heffes.packetized,Sco99,Sco03,Fea06,Lan13}. With the goal of a proper modeling of the doors' failures two-dimensional dataset which results in insight about the reliability of the system, we present in this paper an extension of the $\mmpp$ to the two-dimensional case. The derived stochastic model has the same good analytic properties as the $\mmpp$, namely, a matrix representation, non-exponential distributions of the inter-failures times (and distances), dependence between times and distances, and non-negligible autocorrelation structures for each component. This last property (non-negligible autocorrelation), owned by the analyzed dataset, has been never considered by previous works dealing with bivariate models, to the best of our knowledge. For simplicity, in this paper we will focus on the two-state $\mmpp$, noted as $\mmpp_2$.

	The major contribution of this paper is twofold. On one  hand, we propose an extension of the $\mmpp_2$ to the bivariate case, in such a way that the statistical features previously mentioned are well modeled. Some theoretical properties of this novel process such as identifiability (or existence of unique representation) will also be studied. The second main objective is to propose an estimation approach so that the reliability of the system can be inferred from the observed data. As it will be seen, the approach will be divided into two steps. The first one is based on a matching moments method and provides an estimate for a subset of  parameters, linked to the marginal processes. The second step, which consists in an ABC algorithm, generates estimates for the remaining parameters.


	The paper is structured as follows. Section \ref{sec:studytrains} describes in detail the dataset that motivates the research. Some descriptive summaries are presented to illustrate in clear way the data properties.  After a brief review of the $\mmpp_2$ and the bivariate Marshall-Olkin Exponential distribution in Section \ref{sec:bivprelim}, Section \ref{bivmmpp2} describes the proposed two-dimensional version of the $\mmpp_2$. Section \ref{sec:inden2mmpp} addresses the issue of the identifiability of the process, and Section \ref{sec:bivmatrix} provides a matrix representation for the process. A statistical inference approach for the new bivariate model is considered in Section \ref{fittingbiv}. The performance of the algorithm is illustrated via simulated traces in \ref{simu_biv}. In Section \ref{trenes} a real application of the novel approach is considered to model the dataset related to doors' failures. In the numerical analyses, the estimation of some conditional probabilities of the bivariate processes are also considered. Finally, Section \ref{biv_conclusions} presents conclusions and delineates possible directions for future research.
	
    \section{Case study}\label{sec:studytrains}
	The dataset studied in this paper records failures from 40 underground trains' doors, which were delivered to a European transportation company between November 1989 and March 1991 and all of them were put in service from 20th March 1990 to 20th July 1992. Failure monitoring ended on 31st December 1998. The transportation company was interested in checking the reliability before the expiration of the warranty so that the cost of possible repairs/fixes will be carried by the manufacturer.
	
\begin{figure}[h]
		\centering
		\subfloat[Train 19 ]{{\includegraphics[width=5cm]{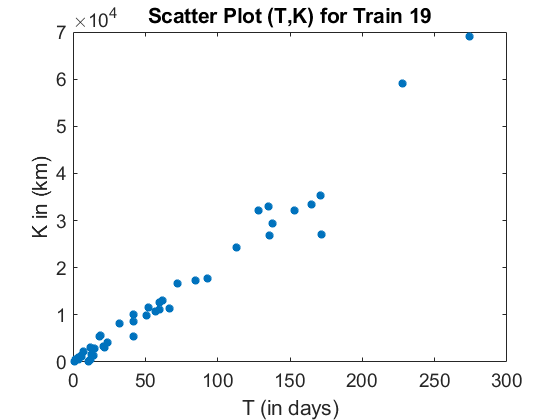}} }
		\subfloat[Train 20]{{\includegraphics[width=5cm]{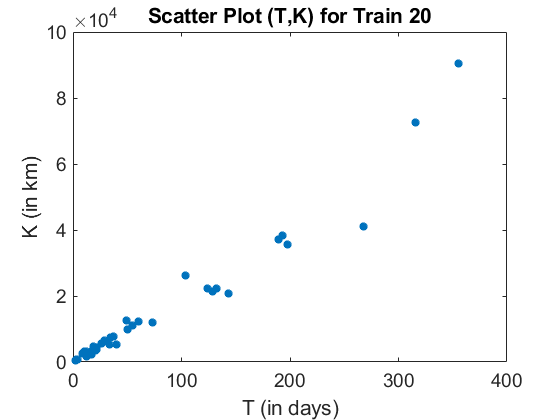} }}%
		
		\subfloat[Train 35]{{\includegraphics[width=5cm]{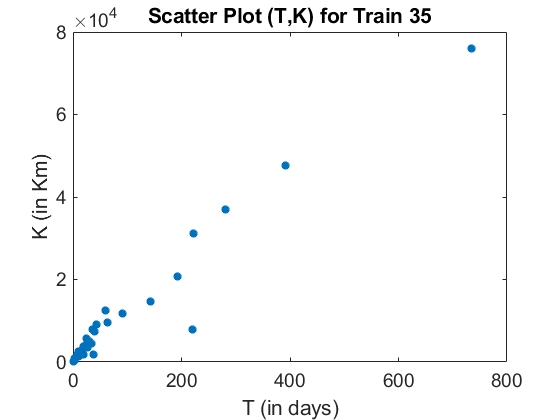} }}%
		\subfloat[Train 36]{{\includegraphics[width=5cm]{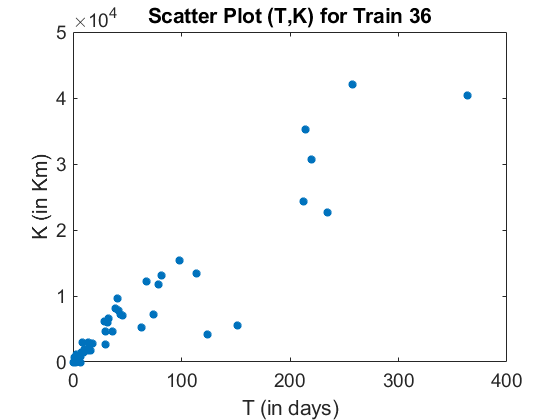} }}%
		\caption{Linear relationship between inter-failure times and distances for four trains}\label{fig:scater}%
	\end{figure}

\begin{figure}[h]
		\centering
		{\includegraphics[width=7cm]{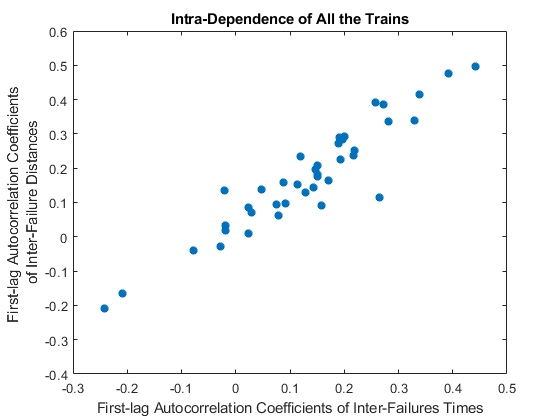}} 
		\caption{Fist-lag autocorrelation coefficient of the inter-failure times and distances for each train}\label{fig:autocorall}%
	\end{figure}

\begin{figure}[h]
		\centering
		\subfloat[Train 35, inter-failure times]{{\includegraphics[width=5cm]{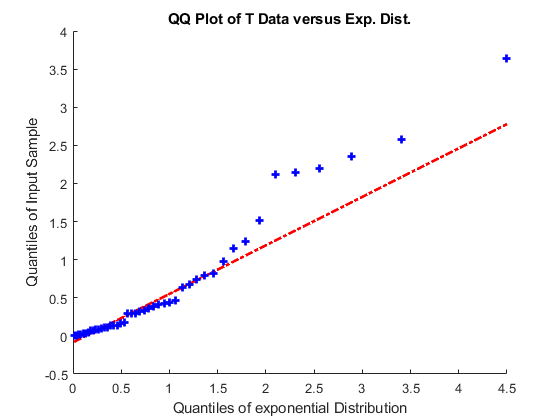}} }
		\subfloat[Train 35, inter-failure distances]{{\includegraphics[width=5cm]{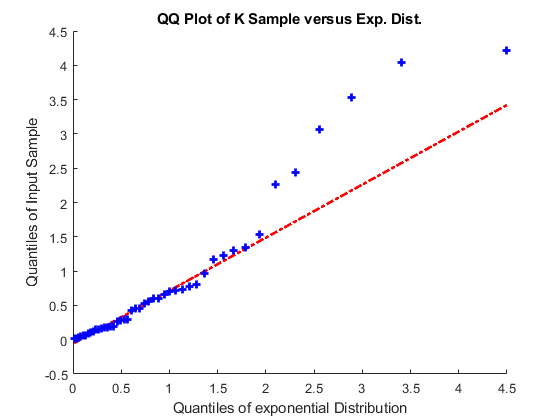} }}%
		
		\subfloat[Train 36, inter-failure times]{{\includegraphics[width=5cm]{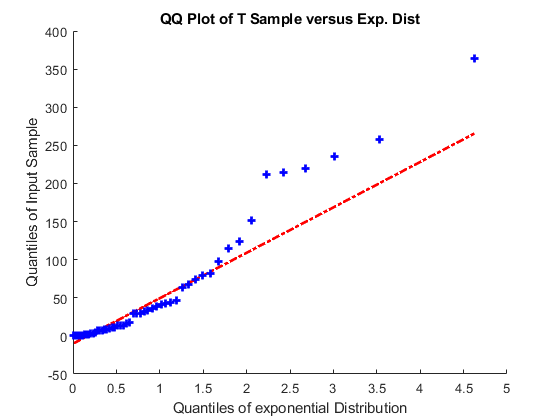} }}%
		\subfloat[Train 36, inter-failure distances]{{\includegraphics[width=5cm]{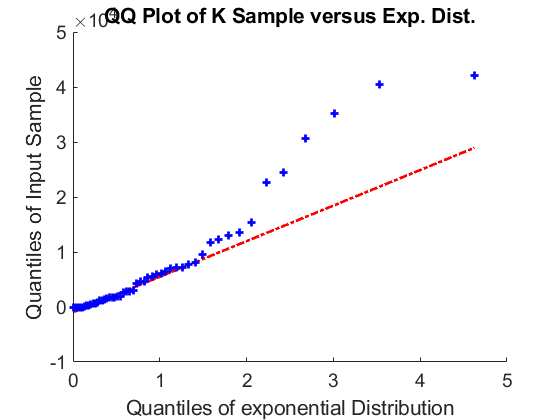} }}%
		\caption{QQ-plots of inter-failure times/distances: empirical versus exponential
			distribution.}\label{fig:qqplotall}%
	\end{figure}

	Figure \ref{fig:scater} shows the scatter plots of inter-failure times and distances covered between the occurrence of two consecutive failures, for four different trains. The figure shows how as the time between failures increases the kilometres travelled by the trains between these failures also increases. It is to be expected that the more time elapses between failures, the more kilometers are covered. In summary, Figure \ref{fig:scater} shows a high linear dependence between the inter-failure times and distances. Throughout this paper the correlation between these two magnitudes will be called inter-dependence.  On the other hand, Figure \ref{fig:autocorall} shows the first-lag autocorrelation coefficients of the inter-failure times versus those of the distances between failures. From the figure, it can be deduced that some autocorrelation coefficients are different from zero for some trains. For those trains it will be said that a non-negligible intra-dependence in the sequence of inter-failures times or distances is observed.  Another important feature of this data is that, neither the inter-failure times, nor the inter-failure distances, seem to be generated by an exponential distribution, as can be figured out from Figure \ref{fig:qqplotall}.

	From the previous properties, it would be desirable to find a bivariate process, versatile enough to allow for non-exponential spacing (both in time and distance) between failures, as well as able to model both intra- and inter-dependence in the observed sequences. We propose in this paper a bivariate version of the $\mmpp_2$ which results in a tractable, analytical model able to jointly fit the dataset's components (time and distance) and from which performance quantities of interest concerning the reliability of the system are derived in straightforward manner. 
	
	It was said in Section \ref{sec:bivintro} that \cite{pievatolo2003} and \cite{pievatolo2010} already analyzed this dataset. In the former \textcolor{black}{a} univariate non-homogeneous Poisson process (NHPP) was considered with an ad hoc relation between time and distance. In the latter paper a bivariate NHPP was considered. Recently, \cite{ruggeri2020} used the same data and a univariate NHPP with 2 parameters to illustrate the properties of their new class of multivariate priors based on stochastic orders. In this paper we consider a bivariate process that, unlike the NHPP, allows for intra-dependence in both the sequences of times and distances. In addition, we do not make any ad hoc assumptions regarding any intensity function. Finally, as it will be seen in Section \ref{trenes} (reliability analysis) we are able to obtain estimates for the joint probability of no failures which, according to the transportation company's purposes, constitutes a measure of crucial importance for the warranty issue.

\section{A bivariate two-state Markov Modulated Poisson Process}\label{sec:mmpp}

	In this section the two-state bivariate Markov Modulated Poisson Process is introduced.   This model can be considered as an extension of the classical two-state Markov Modulated Poisson Process ($\mmpp_2$). \textcolor{black}{This section is divided into four parts.}
The first one is devoted to review the $\mmpp_2$ as well as the Marshall-Olkin bivariate exponential distribution on which the novel bivariate model is based. Section \ref{bivmmpp2} formally introduces the new bivariate \textcolor{black}{process and} an algorithm to simulate traces is described. Section \ref{sec:inden2mmpp} considers the problem of identifiability of the new bivariate model.
\textcolor{black}{Identifiability is an important property in statistical inference ensuring uniqueness of the parameter values for which the optimal value of the maximizing function is obtained.}
Finally, Section \ref{sec:bivmatrix} provides a matrix representation of the process as well as some theoretical quantities of interest.
	
	\subsection{Preliminaries}\label{sec:bivprelim}
	The two-state Markov Modulated Poisson Process, noted $\mmpp_2$, is governed by a two-state underlying Markov process $J(t)$ with infinitesimal generator $\boldsymbol{Q}$ on $\{1,2\}$. 
\textcolor{black}{Then, at the end of an exponentially distributed sojourn time in state $i$, with mean $1/\gamma_i$, two possibilities can occur. First, with probability $a$ if $i=1$ ($b$ if $i=2$), no failure occurs and the system enters into the other state $j \neq i$. Second, with probability $1-a$ if $i=1$ ($1-b$ if $i=2$), a failure is produced and the system continues in the same state.
	The $\mmpp_2$ is also frequently referred to as a Switched Poisson Process ($SPP$), see e.g.~\cite{hoo:83}.}
	
	The $\mmpp_2$ can be characterized in terms of rate (or intensity) matrices  $\{\boldsymbol{D_0},\boldsymbol{D_1}\}$ where
	\begin{equation}
		\boldsymbol{D_0}= \left( \begin{array}{cc}
			-\gamma_1 & \gamma_1 a  \\
			\gamma_2 b & -\gamma_2  \end{array} \right),\quad \boldsymbol{D_1}=\left( \begin{array}{cc}
			\gamma_1 (1-a) & 0  \\
			0 & \gamma_2 (1-b)  \end{array} \right).
	\end{equation}
	This definition of the rate matrices implies that $\boldsymbol{Q}=\boldsymbol{D_0}+\boldsymbol{D_1}$.
	
\begin{figure}[h!]
		\centering
		\includegraphics[scale = .5]{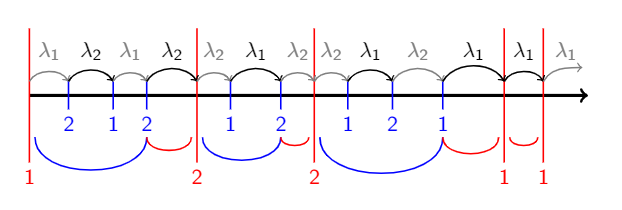}\\
		\caption{Transition diagram for a $\mmpp_2$. The black lines correspond to transitions without failures, governed by $\boldsymbol{D_0}$, and the red lines correspond to transitions with failures, governed by $\boldsymbol{D_{1}}$.}\label{fig:MMPP2}
	\end{figure}

	For a better understanding of the model's behavior, Figure \ref{fig:MMPP2}  illustrates a realization of a $\mmpp_2$. It can be seen \textcolor{black}{how}
\textcolor{black}{states} of the process alternate until the failure takes place. This alternation of states allows to model non-exponential times between failures. Note that the inter-failures \textcolor{black}{times are
a random sum} of exponential times with different failure rates. Since the exponential sojourn times are not commonly observed in real datasets (only the times between failures are), the states of the underlying Markov chain do not have a concrete meaning in practical situations. However, they constitute a mathematical artifact to achieve non-exponentiality and intra-dependence.
	
	Next, some quantities in relation to the reliability of the system are presented. If $T_n$ denotes the time between the $(n-1)$-th and $n$-th failures, then the inter-failure times, $T_n$s, are phase-type distributed  with representation \textcolor{black}{$(\boldsymbol{\phi},\boldsymbol{D_0})$}, where $\boldsymbol{\phi}$ is the stationary probability vector associated with $\boldsymbol{P^*}=\boldsymbol{(-D_0)}^{-1}\boldsymbol{D_1}$, computed as $\boldsymbol{\phi}=(\boldsymbol{\pi} \boldsymbol{D_1} \boldsymbol{e})^{-1}\boldsymbol{\pi} \boldsymbol{D_1}$ (see \cite{Lat90} and \cite{Cha10}), where $\boldsymbol{\pi}$ is the stationary probability vector of $\boldsymbol{Q}$ and $\boldsymbol{e}$ is a vector of ones. This implies that the 	cumulative distribution function of $T_n$ is given by 
	\begin{equation}\label{eq:distMap}
		F_{T_n}(t)=1-\boldsymbol{\alpha_n}e^{\boldsymbol{D_0}t}\boldsymbol{e},
	\end{equation}
	where $\boldsymbol{\alpha_n}=\boldsymbol{\phi}[(-\boldsymbol{D_0})^{-1}\boldsymbol{D_1}]^{(n-1)}=\boldsymbol{\phi}$. Since the process is stationary $\boldsymbol{\phi}$ is defined to satisfy $\boldsymbol{\phi}[(-\boldsymbol{D_0})^{-1}\boldsymbol{D_1}]=\boldsymbol{\phi}$. It is easy to check that for the case of the $\mmpp_2$, $\boldsymbol{\phi}$ only depends on the probabilities associated with the underlying process:
	\begin{equation}\label{eq_phi}
		\boldsymbol{\phi}=(\phi_1,\phi_2)=\left(\frac{b(1-a)}{b(1-a)+a(1-b)},\frac{a(1-b)}{b(1-a)+a(1-b)}\right).
	\end{equation}
The moments of  $T_n$ in the stationary case are given by
	\begin{equation}\label{eq:MAP-mu}
		\mu_T(r) = E (T^r ) = r!\boldsymbol{\phi} (-\boldsymbol{D_0})^{-r}\boldsymbol{e}, \quad \textrm{for } r\geq 1,
	\end{equation}
	and the auto-correlation function of the sequence of inter-failure times is 
	\begin{equation}\label{eq:Tcor}
		\rho_T(l)=\rho (T_1, T_{l+1}) =  \frac{\pi[(-\boldsymbol{D_0})^{-1}\boldsymbol{D_1}]^l(-\boldsymbol{D_0})^{-1}\boldsymbol{e}-\mu_T(1)}{2(\boldsymbol{\pi}(-\boldsymbol{D_0})^{-1}\boldsymbol{e}-\mu_T(1))},\quad \text{for }l>0. 
	\end{equation}
	Further details of the properties of the $\mmpp$ can be found, for example, in \cite{fischer1993}, \cite{ryden961} or \cite{Yera}.

	The construction of the bivariate $\mmpp$ proposed in this article is based on maintaining the same underlying structure of the univariate case, but replacing the exponential univariate distribution with a bivariate exponential distribution. As it is known in the literature, there are several options to define a bivariate exponential distribution and in principle any would serve for the purpose of simulating the bivariate $\mmpp$.
	In this paper we have applied the bivariate Marshall-Olkin exponential distribution, which is a bivariate distribution that fits into a class of Multivariate Phase Type distributions ($MPH^\ast$) proposed in~\cite{kulkarni1989new}, see also Chapter 8 in~\cite{bladt17}.
	The distribution is originally defined in \cite{marshall67} as follows:
	\begin{definition}
		Let $X$ and $Y$ be positive continuous random variables. Then $X$ and $Y$ are distributed according to the bivariate exponential distribution (BVE) with parameters $\lambda_1,\lambda_2,\lambda_3$, noted as $(X,Y)\sim BVE(\lambda_1,\lambda_2,\lambda_3)$ if
		$$\bar{F}(x,y)=P(X>x,Y>y)=\exp\{-\lambda_1 x -\lambda_2 y -\lambda_3\max(x,y) \},$$
		where $\lambda_1,\lambda_2 >0$ and $\lambda_3\geq 0$.
	\end{definition}

	In \cite{marshall67}, basic properties of the BVE are provided. For example, the density function is given by
	\begin{equation*}
		f(x,y)=\left\{\begin{array}{lr}
			\lambda_1 \gamma_k \exp\{-\lambda_1 x -\gamma_k y \} & \textrm{if } x<y  \\
			\lambda_2 \gamma_t \exp\{-\gamma_t x -\lambda_2 y \} & \textrm{if }  x>y\\
			\lambda_3 \exp\{-(\lambda_1 + \lambda_2 + \lambda_3) x \} & \textrm{if }  x=y \end{array} \right.
	\end{equation*}
	The measure has a singular decomposition with a part that is absolutely continuous with respect to the two-dimensional Lebesgue measure in the first quadrant and a measure on the half line $x=y$ in the first quadrant.
	From the definition it can be shown that $X$ and $Y$ follow exponential distributions with means $1/\gamma_t$ and $1/\gamma_k$ respectively, where $\gamma_t=\lambda_1+\lambda_3$ and $\gamma_k=\lambda_2+\lambda_3$.
\cite{marshall67} also deduce the Laplace transform \textcolor{black}{and 
a formula for the joint moments}
	\begin{equation}\label{eq_EXY}
		E(XY)=\frac{1}{\lambda_1+\lambda_2+\lambda_3}\left(\frac{1}{\gamma_t}+\frac{1}{\gamma_k}\right)
	\end{equation}
	and for the correlation between $X$ and $Y$,
	$$\rho(X,Y)=\frac{\lambda_3}{\lambda_1+\lambda_2+\lambda_3}.$$
	The selection of the Marshall-Olkin BVE is also motivated by the fact that the correlation between the two  sequences of  failures  generated by this distribution, $\rho$, is non-negative as it is the case of  the empirical correlation observed in the train failures dataset.
	
	In \cite{kulkarni1989new} and \cite{bladt17} the bivariate Marshall-Olkin exponential distribution is represented as a multivariate phase-type distribution and denoted as $MPH^\ast$. This representation is given by an initial probability vector and two matrices $(\boldsymbol{\alpha},\boldsymbol{S},\boldsymbol{R})$ as follow:
	\begin{equation}\label{eq:mtrix_BVE}
	  \left((1,0,0),\left(\begin{array}{ccc}
	-(\lambda_1 + \lambda_2 + \lambda_3) & \lambda_2 & \lambda_1  \\
	0 & - \gamma_t & 0\\
	0 & 0 & -\gamma_k  \end{array} \right),\left(\begin{array}{cc}
	1 & 1 \\
	1 & 0 \\
	0 & 1  \end{array} \right)\right).  
	\end{equation}
	
	In section \ref{sec:bivmatrix} it will be seen that this  representation of the bivariate Marshall-Olkin exponential distribution will make it possible to obtain a matrix representation for the bivariate $\mmpp_2$. 
	
	\subsection{The bivariate $\mmpp_2$}\label{bivmmpp2}
	
	To formally define the bivariate $\mmpp$, a two-state Markov process $J(t)$ with generator $\boldsymbol{Q}$ on $\{1,2\}$ is considered. When $J(t) = i$, then it is said that the process is in state $i$ and this status remains unchanged while $J(t)$  remains in this state. Specifically, the bivariate $\mmpp_2$ behaves as follows: the initial state $i_0\in \mathcal{S}=\{1,2\}$ is defined according to an initial probability vector \mbox{\boldmath{$\alpha$}}$=(\alpha_1,\alpha_2)$.
The successive bivariate contributions from each state is given by two bivariate Marshall-Olkin exponential distributions. The first one models the contributions from state $i=1$. It is characterized by the parameters  $\boldsymbol{\lambda}=(\lambda_1,\lambda_2,\lambda_3)$, and the marginal distributions have mean $1/\gamma_{t1}$ and $1/\gamma_{k1}$ respectively, with
		\begin{equation}\label{eq:gamma1}
			\gamma_{t1}=\lambda_1+\lambda_3 \quad \textrm{and} \quad \gamma_{k1}=\lambda_2+\lambda_3.
		\end{equation}
		The second BVE associated with state $i=2$ has parameters $\boldsymbol{\omega}=(\omega_1,\omega_2,\omega_3)$ and its marginal exponential distributions have mean $1/\gamma_{t2}$ and $1/\gamma_{k2}$ respectively, with
		\begin{equation}\label{eq:gamma2}
			\gamma_{t2}=\omega_1+\omega_3 \quad \textrm{and} \quad \gamma_{k2}=\omega_2+\omega_3. 
		\end{equation} 

After the increment of the two bivariate variables two possible state transitions can occur. First, with probability $a$ if $i=1$ ($b$ if $i=2$), no failure occurs and the bivariate $\mmpp_2$ enters into the other state $j\neq i$. Second, with probability $1-a$ if $i=1$ ($1-b$ if $i=2$), a failure is produced and the system continues in the same state.

Note that the descriptions of the univariate and bivariate process are very similar. Both have the same underlying process and therefore the same transition probabilities. Essentially, the difference is in the contribution from each visit to a state: while in $\mmpp_2$ two univariate exponential distributions that alternate (one for each state) govern the transitions, in the bivariate case one BVE distribution for each state has to be considered. In other words, \textcolor{black}{the} $\mmpp_2$ is extended to the bivariate case by replacing the exponential distribution by a BVE distribution. Therefore, while the $\mmpp_2$ is determined by 4 parameters, the bivariate $\mmpp_2$ defined in this paper is fully described by 8 parameters; that is, three parameters associated with the Marshal-Olkin BVE distribution in  state $i=1$, $\boldsymbol{\lambda}=(\lambda_1,\lambda_2,\lambda_3)$; another three parameters associated with the Marshal-Olkin BVE distribution in the state $i=2$, $\boldsymbol{\omega}=(\omega_1,\omega_2,\omega_3)$ and the two probabilities associated with the Markov underlying process $(a,b)$.

\begin{figure}[h!]
		\centering
		\includegraphics[scale = .7]{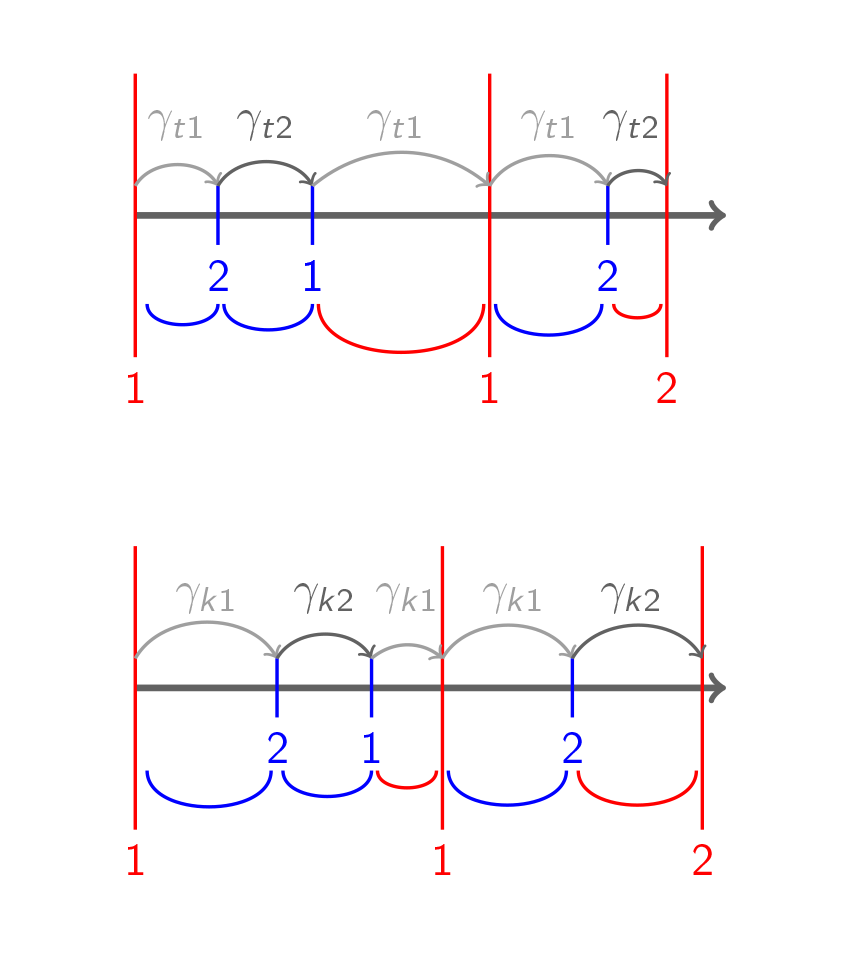}\\
		\caption{A realization of the bivariate $\mmpp_2$. The top panel depicts the sequence of inter-failures times and the bottom panel that of distances covered. The black lines correspond to transitions without failures and the red lines correspond to transitions when a failure occurs.}\label{fig:2MMPP2}
	\end{figure}

Following the notation of the doors’ failures dataset that motivates this research,
we will denote by $\{(T_1,K_1),...,(T_n,K_n)\}$ 
the sample from the bivariate $\mmpp_2$,
where $T$ stands for the time elapsing between two 
consecutive failures and $K$ denotes the
covered distance between such failures. For a better understanding of the considered process, Figure \ref{fig:2MMPP2}  illustrates a realization of the bivariate $\mmpp_2$, where the sequence of inter-failure times ($T_i$) are represented at the top panel while the distances covered ($K_i$) are depicted in the bottom panel. From the figure, it can be seen how both sequences have a common starting point and a common sequence of visited states ($\{1,2,1,1,...\}$).

	For a better clarification of the novel bivariate process, Table \ref{Alg-GenBiv} depicts the algorithm to generate traces from the bivariate $\mmpp_2$.
	
	\begin{table}[h!]
		\begin{center}
			{\tt 
				\begin{enumerate}
					
					\item[0.] Input: \{$a,b,\boldsymbol{\lambda},\boldsymbol{\omega}$\}
					\item[1.] Compute $\boldsymbol{\phi }$ as in (\ref{eq_phi}).
					\item[2.] Generate $p\sim U(0,1)$.
					\item[3.] If $p<\phi_1$, set $s_0=1$ else $s_0=2$.
					\item[4.] Initialize $i=1$, $j=0$, $\boldsymbol{t}=\boldsymbol{0}$ and $\boldsymbol{k}=\boldsymbol{0}$.
					\item[5.] While $i<n$ repeat:
					\begin{enumerate}
						\item[(a)] Generate (X,Y) as a bivariate Marshall-Olkin exponential distribution (BVE). \newline If $s_j=1$, $(X,Y)\sim BVE(\boldsymbol{\lambda})$  else $(X,Y)\sim BVE(\boldsymbol{\omega})$ 
						\item[(b)] Set $(t_i,k_i)=(t_i,k_i)+(X,Y)$ 
						\item[(c)] Generate $p\sim U(0,1)$.
						\item[(d)] if ($s_j=1$ and $p<a$) or ($s_j=2$ and $p<b$), then (there is no failure) $s_{j+1}=|s_{j}-3|$. Else (there is a failure)  $s_{j+1}=s_{j}$ and $i=i+1$.
						\item[(d)] Set $j=j+1$.
					\end{enumerate}
					\item[6.] Output: $\{(t_1,k_1),...,(t_n,k_n)\}$
				\end{enumerate}
			}
		\end{center}
		\caption{\label{Alg-GenBiv} Algorithm to generate a trace from the bivariate $\mmpp_2$}
	\end{table}

		In summary, the bivariate $\mmpp_2$ is defined by two marginal processes sharing the same underlying Markov process and two bivariate Marshall-Olkin exponential distributions. Essentially, the two marginal processes are running alongside, with the same sequence of visited states, the same underlying process and simultaneous failures, but each with different inter-failures "time" rates between failures. The structure of the marginal processes makes it possible to capture the intra-dependence of the data. On the other hand, the existence of a single underlying process relating both marginal processes is what brings about the correlation between the two magnitudes (inter-dependence). 
	
		\subsection{Identifiability of the bivariate \emph{$\mmpp_2$}}\label{sec:inden2mmpp}
	
Identifiability problems occur when different representations of the process lead to the \textcolor{black}{same density functions.
It is known} that $\map$s cannot be identified in a unique way, which is inconvenient for their statistical estimation, see for example \cite{Bod08}, \cite{Ram10} and \cite{Rod16}. On the other hand, \cite{Ryd96} and \cite{Yera} prove the identifiability of the $\mmpp$ and $\bmmpp$. For details about identifiability of the $\mmpp$ and its effect on the estimation, \cite{ryden961}, \cite{Yera} and \cite{yera2019} can be reviewed. 
	
	In this section the identifiability of the proposed bivariate version of the $\mmpp_2$ will be proved. Let us first define the identifiability notion used in this paper.

	\begin{definition}
			Let $B=\{\boldsymbol{\lambda},\boldsymbol{\omega},a,b\}$ be a representation of a bivariate \emph{$\mmpp_2$} and let $T_n$ and $K_n$ denote the sequences of times and distances covered between the ($n-1$)-th and the $n$-th failures. Then, $B$ is said to be identifiable if there exists no different parameterization $\tilde{B}=\{\boldsymbol{\tilde{\lambda}},\boldsymbol{\tilde{\omega}},\tilde{a},\tilde{b}\}$ such that
			\begin{equation}\label{eq_identifiability}
				\{(T_1,K_1),...,(T_n,K_n)\}\stackrel{d}{=}\{(\tilde{T}_1,\tilde{K}_1),...,(\tilde{T}_n,\tilde{K}_n)\} \quad \textrm{for all }n\geq 1,	
			\end{equation}
			where $\tilde{T}_i$ and $\tilde{K}_i$ are defined in analogous way as $T_i$ and $K_i$, and where $\stackrel{d}{=}$ denotes equality in distribution.
	\end{definition}

	\begin{theorem}\label{Theo:BiMMPPInent}
		Let  $B=\{\boldsymbol{\lambda},\boldsymbol{\omega},a,b\}$ and $\tilde{B}=\{\boldsymbol{\tilde{\lambda}},\boldsymbol{\tilde{\omega}},\tilde{a},\tilde{b}\}$ be two different, but equivalent, representations of a bivariate $\mmpp_2$. Then $\boldsymbol{\lambda}={\boldsymbol{\tilde{\lambda}}}$, $\boldsymbol{\omega}={\boldsymbol{\tilde{\omega}}}$ and $(a,b)=(\tilde{a},\tilde{b})$, except for a swap of $a$ by $b$ and $\boldsymbol{\lambda}$ by $\boldsymbol{\omega}$.
	\end{theorem}
	
	See \ref{ap:Theo3} for the proof.

	\subsection{Matrix representation}\label{sec:bivmatrix}
	This section presents one of the most interesting findings regarding the bivariate $\mmpp_2$. It is a matrix representation that eases the analytical and algorithmic results in relation to the quantities of interest associated with the process.
	Indeed, the bivariate $\mmpp_2$ can be represented by $B=\{\boldsymbol{\phi},\boldsymbol{D_0},\boldsymbol{D_1},\boldsymbol{R}\}$, where the initial probability vector is obtained from the stationary probability vector given in (\ref{eq_phi}):
	$$\boldsymbol{\phi}=(\phi_1, 0,0, \phi_2,0, 0)$$
	Note that $\phi_1$ and $\phi_2$, defined in (\ref{eq_phi}), only depend on the probabilities associated with the underlying process. The matrices governing transitions in which failures do not occur and do occur are given by  
	$$\boldsymbol{D_0}=\left( \begin{array}{ccc|ccc}
	-(\lambda_1+\lambda_2+\lambda_3) & \lambda_2 & \lambda_1 & \lambda_3 a & 0 & 0 \\
	0 & -\gamma_{t1} & 0 & \gamma_{t1} a & 0 & 0\\ 
	0 & 0 & -\gamma_{k1} & \gamma_{k1} a & 0 & 0\\
	\hline
	\omega_3 b & 0 & 0 &-(\omega_1+\omega_2+\omega_3) & \omega_2 & \omega_1\\ 
	\gamma_{t2} b& 0 & 0 & 0 &-\gamma_{t2} & 0\\
	\gamma_{k2} b& 0 & 0 & 0 & 0 &-\gamma_{k2}
	\end{array}\right),$$
	and
	$$\boldsymbol{D_1}=\left( \begin{array}{ccc|ccc}
	\lambda_3 (1-a) & 0 & 0 & 0 & 0 & 0 \\
	\gamma_{t1} (1-a) & 0 & 0 & 0 & 0 & 0\\ 
	\gamma_{k1} (1-a) & 0 & 0 & 0 & 0 & 0\\
	\hline
	0 & 0 & 0 &\omega_3 (1-b) & 0 & 0\\ 
	0 & 0 & 0 & \gamma_{t2} (1-b) & 0 & 0\\
	0 & 0 & 0 & \gamma_{k2} (1-b) & 0 & 0
	\end{array}\right)$$
	respectively. Finally
	$$\boldsymbol{R}=\left( \begin{array}{cc}
	1 & 1 \\
	1 & 0 \\ 
	0 & 1 \\
	1 & 1 \\ 
	1 & 0 \\
	0 & 1
	\end{array}\right).$$

\begin{remark}
	Note that the previous matrix representation is not a canonical representation of the process since it is possible to build other matrix representations that fully describe the process (see \ref{ap:MatRepBiv} for an equivalent matrix representation).
The in-depth study of the different matrix representations for this process and the search for a canonical representation is out of the scope of this work and will be addressed in the future.
\end{remark}

\begin{remark}
Note that although matrices $D_0$ and $D_1$ are of dimension $6\times6$, the bivariate $\mmpp_2$ is defined by two states. This is because the sojourn time in each of the two states is modeled through a Marshall-Olkin BVE distribution, whose representation as a $MPH^\ast$ in (\ref{eq:mtrix_BVE}) has three rows. Therefore, while in the matrix representation of the $\mmpp_2$ each state is associated with a column or a row, in the bivariate $\mmpp_2$  there are three rows or columns associated with each state of the process.
\end{remark}
	
From the matrix representation $B=\{\boldsymbol{\phi},\boldsymbol{D_0},\boldsymbol{D_1},\boldsymbol{R}\}$, it can be deduced that $(T_n,K_n)$ follows a multivariate phase type distribution $MPH^\ast$ as defined in \cite{kulkarni1989new} and \cite{bladt17}, with representation $\{\boldsymbol{\phi},\boldsymbol{D_0},\boldsymbol{R}\}$. This result allows for the use of Theorem 8.1.2. in \cite{bladt17} for obtaining the moment-generating function until the occurrence of the first failure in the  bivariate $\mmpp_2$. We recall that $\Delta(a)$ denotes the diagonal matrix with vector $a$ as diagonal.

	\begin{proposition}\label{Theo:FunGenMom}
		Let $B=\{\boldsymbol{\phi},\boldsymbol{D_0},\boldsymbol{D_1},\boldsymbol{R}\}$ be a representation of a bivariate\emph{ $\mmpp_2$} and let $\boldsymbol{A}=(T_1,K_1)$ be the records related to the first failure. Then there exists a $K>0$ such that the moment-generating function for $\boldsymbol{A}$ (denoted by  $H(\boldsymbol{\theta})$)
		exists and is given by
		$$H(\boldsymbol{\theta})=E\left(e^{\boldsymbol{A}\boldsymbol{\theta}}\right)=\boldsymbol{\phi}\left(-\boldsymbol{\Delta}(\boldsymbol{R}\boldsymbol{\theta})-\boldsymbol{D_0}\right)^{-1}\boldsymbol{D_1}\boldsymbol{e},$$
		for any \textcolor{black}{$\theta_1,\theta_2<K$ with} $\boldsymbol{\theta}=[\theta_1,\theta_2]^t$.
	\end{proposition}
	
\begin{proof}
	Since $A\sim MPH^\ast(\phinegrita,\boldsymbol{D_0},\boldsymbol{R})$, then Theorem 8.1.2. in \cite{bladt17} can be applied directly. Substituting $\boldsymbol{\alpha}=\boldsymbol{\phi}$, $\boldsymbol{S}=\boldsymbol{D_0}$ and $\boldsymbol{s}=\boldsymbol{D_1}\boldsymbol{e}$ the result is immediate.
\end{proof}

	From Proposition \ref{Theo:FunGenMom}, the formulae for moments and cross moments of the process, which are of interest when studying the reliability of the system, are obtained as follows.

	\begin{proposition}\label{Theo:BiMMPPMom}
		Let $B=\{\boldsymbol{\phi},\boldsymbol{D_0},\boldsymbol{D_1},\boldsymbol{R}\}$ be a representation of a bivariate \emph{$\mmpp_2$}. Then, the joint moments of $(T_i, K_i)$ are given by 
		\begin{equation}\label{eq:ETnKm}
			\eta_{nm}=E\left(T_i^nK_i^m\right)=\boldsymbol{\phi}\sum_{i=1}^{(m+n)!}\left(\prod_{j=1}^{n+m}(-\boldsymbol{D_0})^{-1}\boldsymbol{\Delta}(\boldsymbol{R}_{\cdot\sigma_i(j)})\right)\boldsymbol{e},
		\end{equation}
		where $\boldsymbol{R_{\cdot j}}$ is $j^{th}$ column of $\boldsymbol{R}$, $\sigma_1, . . . ,\sigma_{(n+m)!}$ are the ordered permutations of duples of derivatives, and $\sigma_i(j)\in\{1,2\}$ is the $i^{th}$ position of the permutation $\sigma_i$.
	\end{proposition}
	
	\begin{proof}
As the pair of the variables $(T_i,K_i)$ follows a multivariate phase type distribution $MPH^\ast$, Theorem 8.1.5 in \cite{bladt17} can be applied directly to $({T_i,K_i})$. Substituting $\boldsymbol{\alpha}=\boldsymbol{\phi}$, and $\boldsymbol{U}=\boldsymbol{(-D_0)}^{-1}$ the result is obtained.
	\end{proof}
	
	Some particular joint moments used in Section \ref{fittingbiv} for designing the estimation approach for the bivariate $\mmpp_2$ are:
	\begin{eqnarray}\label{eq:mometa}
			\eta_{11}=E(TK)&=&\boldsymbol{\phi}(-\boldsymbol{D_0})^{-1}\boldsymbol{\Delta}(\boldsymbol{R_{[\cdot 1]}})(-\boldsymbol{D_0})^{-1}\boldsymbol{R_{[\cdot 2]}} \\
		& & +\boldsymbol{\phi}(-\boldsymbol{D_0})^{-1}\boldsymbol{\Delta}(\boldsymbol{R_{[\cdot 2]}})(-\boldsymbol{D_0})^{-1}\boldsymbol{R_{[\cdot 1]}}\nonumber\\
		\eta_{21}=E(T^2K)&=&2\boldsymbol{\phi}(-\boldsymbol{D_0})^{-1}\boldsymbol{\Delta}(\boldsymbol{R_{[.1]}})(-\boldsymbol{D_0})^{-1}\boldsymbol{\Delta}(\boldsymbol{R_{[.1]}})(-\boldsymbol{D_0})^{-1}\boldsymbol{R_{[.2]}}\nonumber\\
		&&+2\boldsymbol{\phi}(-\boldsymbol{D_0})^{-1}\boldsymbol{\Delta}(\boldsymbol{R_{[.1]}})(-\boldsymbol{D_0})^{-1}\boldsymbol{\Delta}(\boldsymbol{R_{[.2]}})(-\boldsymbol{D_0})^{-1}\boldsymbol{R_{[.1]}}\nonumber\\
		&&+2\boldsymbol{\phi}(-\boldsymbol{D_0})^{-1}\boldsymbol{\Delta}(\boldsymbol{R_{[.2]}})(-\boldsymbol{D_0})^{-1}\boldsymbol{\Delta}(\boldsymbol{R_{[.1]}})(-\boldsymbol{D_0})^{-1}\boldsymbol{R_{[.1]}}\nonumber \\
		\eta_{12}=E(TK^2)&=&2\boldsymbol{\phi}(-\boldsymbol{D_0})^{-1}\boldsymbol{\Delta}(\boldsymbol{R_{[.2]}})(-\boldsymbol{D_0})^{-1}\boldsymbol{\Delta}(\boldsymbol{R_{[.2]}})(-\boldsymbol{D_0})^{-1}\boldsymbol{R_{[.1]}}\nonumber\\
		&&+2\boldsymbol{\phi}(-\boldsymbol{D_0})^{-1}\boldsymbol{\Delta}(\boldsymbol{R_{[.2]}})(-\boldsymbol{D_0})^{-1}\boldsymbol{\Delta}(\boldsymbol{R_{[.1]}})(-\boldsymbol{D_0})^{-1}\boldsymbol{R_{[.2]}}\nonumber\\
		&&+2\boldsymbol{\phi}(-\boldsymbol{D_0})^{-1}\boldsymbol{\Delta}(\boldsymbol{R_{[.1]}})(-\boldsymbol{D_0})^{-2}\boldsymbol{\Delta}(\boldsymbol{R_{[.1]}})(-\boldsymbol{D_0})^{-1}\boldsymbol{R_{[.2]}}\nonumber,
	\end{eqnarray}
	where $\boldsymbol{R_{[.j]}}$ is the $j^{th}$ column of $\boldsymbol{R}$.\bigskip

\begin{figure}[h!]
		\centering
		\subfloat[$corr(T,K)=0.38$ ]{{\includegraphics[width=5cm]{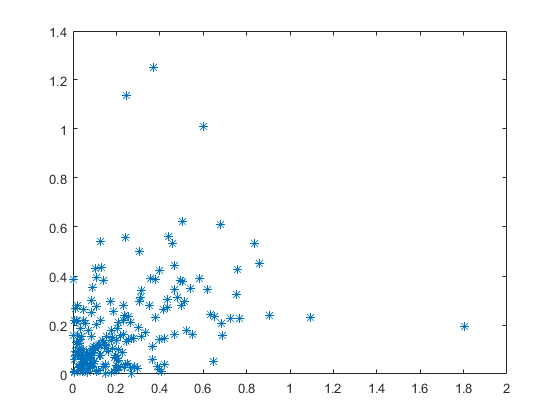}}}
		\subfloat[$corr(T,K)=0.64$]{{\includegraphics[width=5cm]{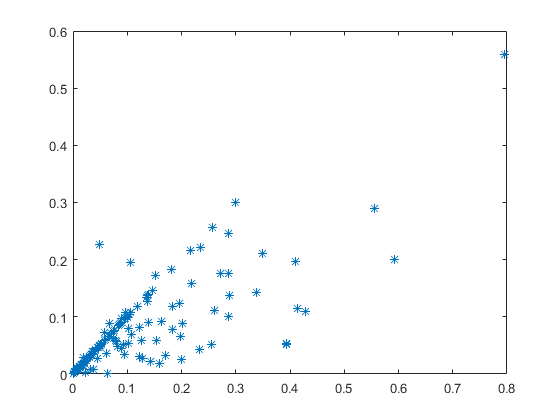} }}%
		
		\subfloat[$corr(T,K)=0.84$]{{\includegraphics[width=5cm]{plot_sim_084.png} }}%
		\subfloat[$corr(T,K)=0.94$]{{\includegraphics[width=5cm]{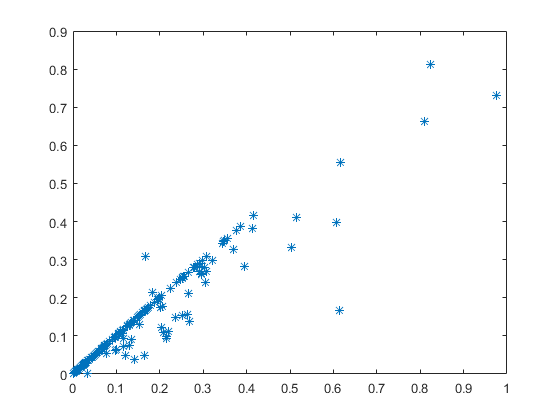} }}%
		\caption{Scatter plots of simulated sequences generated from bivariate $\mmpp_2$s.}\label{fig:bivcorr}
	\end{figure}

	From the expressions previously obtained for $\eta_{11}$, $\eta_{21}$ and $\eta_{12}$ it is not difficult to deduce a closed expression for the correlation coefficient between the sequences of times and distances, $corr(T,K)$. In order to show the versatility of the bivariate $\mmpp_2$, Figure \ref{fig:bivcorr} depicts four scatter plots of simulated sequences for different representations. \bigskip


	\textcolor{black}{The next result} generalizes Proposition \ref{Theo:FunGenMom} and gives the moment generating function of the first $n$ failures.

	\begin{proposition}\label{col:Transformada}
		The moment generating function 
		of the $n$ first consecutive failures $\{(T_1, K_1), (T_2,K_2),\ldots, (T_n,K_n)\}$ is given by
		\begin{eqnarray*}
		f^*_{\{(T_1,K_1),...,(T_n,K_n)\}}(\boldsymbol{\theta_1},...,\boldsymbol{\theta_n}) & = & E\left(e^{-\sum_{i=1}^n(T_i,K_i)\left(\begin{array}{c}\theta_{i1} \\ \theta_{i2} \end{array}\right)}\right)\\
		& = & \boldsymbol{\phi} (-\boldsymbol{\Delta}(\boldsymbol{R}\boldsymbol{\theta}_1)-\boldsymbol{D}_0)^{-1}\boldsymbol{D}_1 \dots(-\boldsymbol{\Delta}(\boldsymbol{R}\boldsymbol{\theta}_n)-\boldsymbol{D}_0)^{-1}\boldsymbol{D}_1\boldsymbol{e}
		\end{eqnarray*}
	\end{proposition}

	\begin{corollary}\label{col:seqcor}
		With $X_1=T_1$ or $X_1=S_1$ and $Y_{1+n}=T_{1+n}$ or $Y_{1+n}=S_{1+n}$ we have
		\begin{eqnarray*}
			E(X_1Y_{1+n}) & =  &\boldsymbol{\phi}(-\boldsymbol{D_0})^{-1}\boldsymbol{R_{[\cdot i]}}\boldsymbol{P}^{n-1}(-\boldsymbol{D_0})^{-1}\boldsymbol{R_{[\cdot j]}}\boldsymbol{e}  ,\quad n=1,2\dotsc
		\end{eqnarray*}
where $\boldsymbol{P} = (-\boldsymbol{D_0})^{-1}\boldsymbol{D_1}$, $i=1$ for $X_1=T_1$, $i=2$ for $X_1=S_1,$ $j=1$ for $Y_{1+n}=T_{1+n}$, and $j=2$ for $Y_{1+n}=S_{1+n}.$
	\end{corollary}

		\begin{remark}
		From Proposition \ref{col:Transformada} it can be proved that the marginal processes of the bivariate \emph{$\mmpp_2$} are univariate \emph{$\mmpp_2$}s represented by the following rate matrices:
		\begin{equation}\label{eq_M1}
			B_t=\left\{\boldsymbol{D_{0t}}=\left( \begin{array}{cc}
				-\gamma_{t1} & \gamma_{t1}a  \\
				\gamma_{t2}b & -\gamma_{t2}  \end{array} \right), \quad \boldsymbol{D_{1t}}=\left( \begin{array}{cc}
				\gamma_{t1} (1-a) & 0  \\
				0 & \gamma_{t2}(1-b)  \end{array} \right)\right\}
		\end{equation}
		and
		\begin{equation}\label{eq_M2}
			B_k=\left\{\boldsymbol{D_{0k}}=\left( \begin{array}{cc}
				-\gamma_{k1} & \gamma_{k1}a  \\
				\gamma_{k2}b & -\gamma_{k2}  \end{array} \right), \quad \boldsymbol{D_{1k}}=\left( \begin{array}{cc}
				\gamma_{k1}(1-a) & 0  \\
				0 & \gamma_{k2}(1-b)  \end{array} \right)\right\}.
		\end{equation}
	\end{remark}

		\begin{remark}
Proposition~\ref{col:Transformada} and Corollary~\ref{col:seqcor} have been formulated for the bivariate \emph{$\mmpp_2$} process, but \textcolor{black}{they hold} in the general case of a sequence of correlated $MPH^\ast$ variables.  In such case, one needs to consider $\boldsymbol{\theta}$ vectors of higher dimension \textcolor{black}{and make proper substitutions of the random variables.}
Similarly, Corollary~\ref{col:seqcor} can be extended so that higher powers are considered and more variables are included.
	\end{remark}

	\section{Inference for the bivariate $\mmpp_2$}\label{fittingbiv}
	In this section, an approach for estimating the parameters of the bivariate $\mmpp_2$ is proposed. Since, in practice, the complete sequence of visited states of the underlying Markov process are not observed, the proposed procedure assumes that $\boldsymbol{t} = (t_1, t_2, . . . , t_n)$, and  $\boldsymbol{k} = (k_1, k_2, . . . , k_n)$, are the only available information. An important issue to take into account is the complication when applying the likelihood principle to a singular measure. 
Therefore, the proposed fitting algorithm avoids evaluating or optimizing the likelihood. Indeed, the algorithm is a two-step approach. In the first one, the rate matrices of the marginal processes $\boldsymbol{D_{0t}}, \boldsymbol{D_{1t}}, \boldsymbol{D_{0k}}, \boldsymbol{D_{1k}}$ are estimated through a moments matching method defined by a standard optimization problem. The remaining parameters are estimated in a second step via an ABC algorithm. 

	
	\cite{Bod08} characterizes the $\mmpp_2$ by the first three moments and first-lag auto-correlation coefficient 
		of the inter-failure time distribution,
	$\mu_T(1), \mu_T(2), \mu_T(3)$ to the one in (\ref{eq:MAP-mu}) and $\rho_T(1)$ as in (\ref{eq:Tcor}). This implies that, for the bivariate $\mmpp_2$ considered in this paper, the matrix representation of the marginal processes as in (\ref{eq_M1}) and (\ref{eq_M2}) will be characterized by a set of eight moments. Four of them are related to the  variable ($T$), \{$\mu_T(1)$, $\mu_T(2)$, $\mu_T(3)$, $\rho_T(1)$\} and the other four to the variable ($K$), $\{\mu_K(1)$, $\mu_K(2)$, $\mu_K(3)$, $\rho_K(1)$\}. 
	
\cite{Carpepa}, using the results found by \cite{Bod08}, derive a moments matching method for estimating the parameters of a $\map_2$, given a sequence of inter-failure times $\boldsymbol{t} = (t_1, t_2, . . . , t_n)$. Posterior adaptations of this procedure can be found in \cite{Rod161} and \cite{Yera} to estimate the non-stationary $\map$ and $\bmmpp_2$ respectively. In the case of the bivariate $\mmpp_2$, it is an open problem to represent the process by a set of moments. However, a moments matching approach similar to the one in \cite{Carpepa} can be designed to partially estimate its parameters. In particular, the parameters associated with the marginal processes $\boldsymbol{D_{0t}}$, $\boldsymbol{D_{1t}}$, $\boldsymbol{D_{0k}}$ and $\boldsymbol{D_{1k}}$ in terms of $\gamma_{t1},\gamma_{t2},\gamma_{k1},\gamma_{k2}, a$ and $b$ (see (\ref{eq_M1}) and (\ref{eq_M2})), can be estimated as the solution of the following optimization problem:
	
	\begin{equation*}(P0)
		\left\{
		\begin{array}{lll}
			\underset{\boldsymbol{\gamma}, a,b}{\min} & \displaystyle \delta_{0}\color{black}(a, b,\gamma_{t1},\gamma_{t2},\gamma_{k1},\gamma_{k2}) \\
			\mbox{s.t.} &  \gamma_{t1},\gamma_{t2},\gamma_{k1},\gamma_{k2} \geq 0,\\
			& 0 \leq a, b \leq 1,\\
		\end{array}
		\right.
	\end{equation*}
	The objective function in (P0) is
	\begin{equation*}
		\begin{array}{rcl}	\delta_{0}(\gamma_{t1},\gamma_{t2},\gamma_{k1},\gamma_{k2}, a, b) &=& [\rho_T(1)(a,b,\gamma_{t1},\gamma_{t2})-\bar{\rho}_T(1)]^2\\ 
			&&+ [\rho_K(1)(a,b,\gamma_{k1},\gamma_{k2})-\bar{\rho}_K(1) ]^2 \\
			&&
			+\sum_{j=1}^{3}\left(\frac{\mu_T(j)(a,b,\gamma_{t1},\gamma_{t2})-\bar{\mu}_T(j)}{\bar{\mu}_T(j)}\right)^2\\
			&&+\sum_{j=1}^{3}\left(\frac{\mu_K(j)(a,b,\gamma_{k1},\gamma_{k2})-\bar{\mu}_K(j)}{\bar{\mu}_K(j)}\right)^2,
		\end{array}
\end{equation*}
where $\bar{\mu}_T(i)$, for $i=1,2,3$ and $\bar{\rho}_T(1)$  denote the empirical moments associated with the first marginal process (computed from the sample $\boldsymbol{t}$), whereas $\bar{\mu}_K(i)$, for $i=1,2,3$ and $\bar{\rho}_K(1)$  denote the empirical moments associated with the second one (computed from the sample $\boldsymbol{k}$).
Note that in the previous objective function,  $\rho_T(1)$ and $\mu_T(i)$, for $i=1,2,3$ depend on $a,b,\gamma_{t1}$ and $\gamma_{t2}$, while $\rho_K(1)$ and $\mu_K(i)$ for $i=1,2,3$ depend on $a, b,\gamma_{k1}$ and $\gamma_{k2}$.
Probabilities $a$ and $b$ are common for the two marginal process since both marginal $\mmpp_2$s share the same underlying Markov process.
For more details of this procedure, see the algorithm of Table 2.

		\begin{table}[H]
		\begin{center}
			{\tt 
				\begin{enumerate}
					\item[0.] Input: \{$\bar{\mu}_T(1),\bar{\mu}_T(2),\bar{\mu}_T(3),\bar{\rho}_T(1),\bar{\mu}_K(1),\bar{\mu}_K(1),\bar{\mu}_K(1),\bar{\rho}_K(1)$\}
					\item[1.] For $i=1,\ldots,I$ repeat:
					\begin{enumerate}
						\item[(a)] Randomly select a starting point $\{a^{(i)}(0),b^{(i)}(0),\gamma_{t1}^{(i)}(0),\gamma_{t2}^{(i)}(0),\gamma_{k1}^{(i)}(0),\gamma_{k2}^{(i)}(0) \}$. 
						\item[(b)] Solve $(P0)_i$ and save the value of objective function $\delta_{0}^{(i)}$ and the solution $\{\hat{a}^{(i)}, \hat{b}^{(i)},\widehat{\gamma}_{t1}^{(i)},\widehat{\gamma}_{k1}^{(i)},\widehat{\gamma}_{t2}^{(i)},\widehat{\gamma}_{k2}^{(i)} \}$.  
					\end{enumerate}
					\item[2.] Obtain $j=\arg\min_{i}\delta_{0}^{(i)}$ and set $\delta_{0}=\delta_{0}^{(j)}, \widehat{\gamma}_{t1}=\widehat{\gamma}_{t1}^{(j)}, \widehat{\gamma}_{t2}=\widehat{\gamma}_{t2}^{(j)}, \widehat{\gamma}_{k1}=\widehat{\gamma}_{k1}^{(j)}, \widehat{\gamma}_{k2}=\widehat{\gamma}_{k2}^{(j)}, \hat{a}=\hat{a}^{(j)}$ and $\hat{b}=\hat{b}^{(j)}$
					\item[3.] Output: $\{  \hat{a}, \hat{b},\widehat{\gamma}_{t1},\widehat{\gamma}_{t2},\widehat{\gamma}_{k1},\widehat{\gamma}_{k2}\}$
				\end{enumerate}
			}
		\end{center}
		\caption{\label{Alg-1}Step 1 in the fitting approach: moments matching method to estimate the parameters of the marginal components in the bivariate $\mmpp_2$}
	\end{table}
	
Note that we propose to solve problem (P0) a number $I$ of times, so that the final solution will be the one that provides the lowest objective function.
Each time, problem (P0) is solved using a different starting point $(a(0),b(0),\gamma_{t1}(0),\gamma_{t2}(0),\gamma_{k1}(0),\gamma_{k2}(0))$ to avoid getting stuck at a poor local optimum.
In practice, $I$ is set to be equal to 100, which has proven in the numerical experiments to be high enough so as to get reasonable solutions.
The final solution $\{\hat{a},\hat{b},\hat{\gamma}_{t1},\hat{\gamma}_{t2},\hat{\gamma}_{k1},\hat{\gamma}_{k2}\}$ shall be considered the starting point for the second step in the proposed fitting algorithm, an ABC approach whose output is the complete set of parameters characterizing the bivariate 	$\mmpp_2$, that is $\{a,b,\boldsymbol{\lambda},\boldsymbol{\omega}\}$.
From the algorithm in Table 2, the estimated values of $a$ and $b$ are obtained.
With regards to the values of $\boldsymbol{\lambda}$ and $\boldsymbol{\omega}$ note that since 	$\gamma_{t1}=\lambda_1+\lambda_3$, $\gamma_{t2}=\omega_1+\omega_3$, $\gamma_{k1}=\lambda_2+\lambda_3$ and  $\gamma_{k2}=\omega_2+\omega_3$, see (\ref{eq:gamma1}) and (\ref{eq:gamma2}), the estimated values for $\lambda_1+\lambda_3$, $\lambda_2+\lambda_3$, 	$\omega_1+\omega_2$ and $\omega_2+\omega_3$ are obtained from the moment matching approach.
On the other hand, as it has been mentioned previously, it is not trivial to evaluate the likelihood function when a singular measure is involved. On the contrary, it is very straightforward to generate simulated traces from the bivariate $\mmpp_2$, see the algorithm in Table 1. In this setting, the ABC algorithm turns out suitable, see for example  \cite{csillery2010}, \cite{marin2012} and \cite{kypraios2017}. The ABC algorithm is mathematically well-founded  and  applied in a wide variety of fields, but there are some issues that have to be carefully considered for a good performance, as the dimension of the parameter space. The larger the dimension of the parameter space is, the more simulations are needed since, as \cite{csillery2010} points out, the probability of accepting the simulated values for the parameters under a given tolerance decreases exponentially when increasing dimensionality. 

	In the case of the bivariate $\mmpp_2$, estimates for $a$, $b$, $\lambda_1+\lambda_3$ ($\gamma_{t1}$), $\lambda_2+\lambda_3$ ($\gamma_{k1}$), $\omega_{1}+\omega_{3}$ ($\gamma_{t2}$) and $\omega_{2}+\omega_{3}$ ($\gamma_{k2}$) are obtained according to the algorithm of Table 2. Taking into account expressions (\ref{eq:gamma1}) and (\ref{eq:gamma2}) and setting prior distributions for
		$\lambda_3$ and $\omega_{3}$ as
		\begin{equation}\label{eq:lam3ome3}
			\lambda_3 \sim Unif(0,\min[\widehat{\gamma}_{t1},\widehat{\gamma}_{k1}]), \quad
			\omega_3 \sim Unif(0,\min[\widehat{\gamma}_{t2},\widehat{\gamma}_{k2}]), 
		\end{equation}
		then, a simple ABC algorithm in terms of only two parameters can be easily formulated. At each iteration $i\in{1,…,I}$, values of $\lambda_3$ and $\omega_3$ are generated from the prior $\pi(\cdot)$ as in (\ref{eq:lam3ome3}). Then, the values of $\lambda_1$, $\lambda_2$, $\omega_1$ and $\omega_2$ are obtained according to (\ref{eq:gamma1}) and (\ref{eq:gamma2}). 
		Then, a sample $\boldsymbol{s}^{(i)}=\{(t_1,k_1),…,(t_n,k_n)\}$ from a bivariate $\mmpp_2$ with parameters $(\hat{a},\hat{b},\boldsymbol{\lambda}^{(i)},\boldsymbol{\omega}^{(i)})$ is simulated according to the algorithm of Table 1. If the generated sample $\boldsymbol{s}^{(i)}$ is too different from the observed data $\boldsymbol{s}=\{(t_1,k_1),...,(t_n,k_n)\}$, the parameter set is discarded.  For this purpose a distance measure and a tolerance, $\epsilon>0$, are usually established. The level of discrepancy between the generated sample at iteration $i$ and the original sample
		shall be measured according to
	\begin{equation}\label{eq:distance}
		\delta_{1}(s^{(i)},s) =  \sum_{l=1,j=1}^{2}\left(\frac{\bar{\eta}_{lj}(s^{(i)})-\bar{\eta}_{lj}(s)}{\bar{\eta}_{lj}(s)}\right)^2,
	\end{equation}
	where $\bar{\eta}_{lj}$, for $l,j=1,2$ denote the first empirical joint moments associated with the bivariate $\mmpp_2$ process ($E(TK),E(T^2K), E(TK^2)$) as in (\ref{eq:mometa}). The rationale for this choice is the fact that for all iterations, the simulated samples come from a bivariate $\mmpp_2$ processes with the same marginal moments (a consequence of constant $\hat{\gamma}_{t1},\hat{\gamma}_{t2},\hat{\gamma}_{k1}, \hat{\gamma}_{t1}, \hat{a}$ and $\hat{b}$). Therefore, it makes sense to include joint moments in the distance measure. From extensive, empirical experiments it has been observed that $\eta_{11}, \eta_{12}$ and $\eta_{21}$ in combination with the eight marginal moments used in the previous moments matching approach $(\mu_T(1),\mu_T(2),\mu_T(3),\rho_T(1),\mu_K(1),\mu_K(2),\mu_K(3)\rho_K(1))$ are enough to characterize the parameters of the bivariate process. With respect to the tolerance level, instead of fixing a specific value of $\epsilon$, we proceed in analogous way by keeping the $1\%$ of the samples with smallest differences from the original sample. Finally, the estimated parameters are average values among the selected proportion. For a summary of the ABC procedure, see the algorithm of Table 3. The performance of the method has been tested through simulated and real datasets. For a detailed analysis concerning simulated data, we refer the reader to \ref{simu_biv}.
	\begin{table}[t]
		\begin{center}
			{\tt 
				\begin{enumerate}
					\item[0.] Input: \{$\bar{\eta}_{11},\bar{\eta}_{12},\bar{\eta}_{21},\hat{a}, \hat{b},\widehat{\gamma}_{t1},\widehat{\gamma}_{k1},\widehat{\gamma}_{t2},\widehat{\gamma}_{k2}$\}
					\item[1.] For $i=1,\ldots,I_2$ do repeat:
					\begin{enumerate}
						\item[(a)] Generate  $(\lambda_3^{(i)},\omega_3^{(i)})$
						from the prior distribution $\pi(\cdot)$
						\item[(b)] Obtain
						\begin{eqnarray*}
							\lambda_1^{(i)}&=&\hat{\gamma}_{t1}-\lambda_3^{(i)}\\
							\lambda_2^{(i)}&=&\hat{\gamma}_{k1}-\lambda_3^{(i)}\\
							\omega_2^{(i)}&=&\hat{\gamma}_{t2}-\omega_3^{(i)}\\
							\omega_2^{(i)}&=&\hat{\gamma}_{k2}-\omega_3^{(i)}
						\end{eqnarray*} 
						\item[(c)] Simulate a sample $\boldsymbol{s}^{(i)}$ from the likelihood \mbox{$f(\cdot \mid \hat{a},\hat{b},\boldsymbol{\lambda}^{(i)},\boldsymbol{\omega}^{(i)})$}.
						\item[(c)] Compute the moments $\bar{\eta}_{11}, \bar{\eta}_{12}, \bar{\eta}_{21}$  associated with the generated sample $\boldsymbol{s}^{(i)}$.
						\item[(d)] Compute $\delta_{1}^{(i)}(\boldsymbol{s}^{(i)},\boldsymbol{s})$ as in (\ref{eq:distance})
					\end{enumerate}
					\item[4.] The 1\% of the sampled values with the smallest differences from the real data are accepted.
					\item[5.] The Bayesian estimates are computed as the average of the accepted values.
					\item[6.] Output: $\{\hat{a}, \hat{b},\hat{\lambda}_1,\hat{\lambda}_2,\hat{\lambda}_3,\hat{\omega}_1,\hat{\omega}_2,\hat{\omega}_3\}$
				\end{enumerate}
			}
		\end{center}
		\caption{\label{Alg-2}Step 2 in the fitting approach: ABC method to estimate the parameters of the bivariate $\mmpp_2$}
	\end{table}

	\section{Reliability analysis of  train failures}\label{trenes}
	
	In this section we address the motivating example for the paper, showing how the stochastic model introduced in earlier sections can be used to analyze the two-dimensional data about train failures and get insights about the reliability of the system. 
	
	In Section \ref{Performance} the fit of the bivariate $\mmpp_2$ model is illustrated. The moments characterizing the process as well as the cumulative distribution functions for the components (inter-failure and distances) are estimated. Once the good performance of the bivariate $\mmpp_2$ is proven, then Section \ref{reliability_data} deals with the estimation of key quantities concerning the reliability of the system. 
	
	\subsection{Performance of the fitting approach}\label{Performance}
	Consider the real dataset described in Section \ref{sec:studytrains}. Train 35 went into operation on 20/12/90 and, in the almost 8 years, it traveled 359908 km and suffered 47 failures. The  average, median, variation coefficient, minimum and maximum value of the inter-failures times for train 35 are $71.12$, $24$, $1.92$, $1$ and $736$ days, respectively, which suggest a right-skewed distribution with a tail longer than that of an exponential distribution, a fact also deduced from Figure \ref{fig:qqplotall}. Something similar occurs with the inter-failures distances, with the average, median, variation coefficient, minimum and maximum equal to $8778$, $3592$, $1.6842$, $87$ and $75998$ kilometers. On the other hand, for train 36 there are records between 04/09/90 and 04/09/98, period in which it covered 379709 km and whose doors failed 51 times. The analysis carried out on the non-exponentiality of the inter-failures times and distances for train 35 applies also for train 36. Empirically, the consecutive inter-failure times and distances are not independent since the first lag autocorrelation coefficients are equal to 0.13 and 0.21, for trains 35 and 36, respectively. There is also inter-dependence among the traces, reflected by an empirical correlation coefficient equal to 0.97 and 0.93 for trains 35 and 36, respectively. The non-exponentiality of the traces in combination with both intra- and inter-dependence makes the bivariate $\mmpp_2$ proposed in this paper a suitable model for fitting the data.
	
\begin{figure}[H]
		\begin{center}
			\begin{tabular}{c c}
				\includegraphics[height=1.7in]{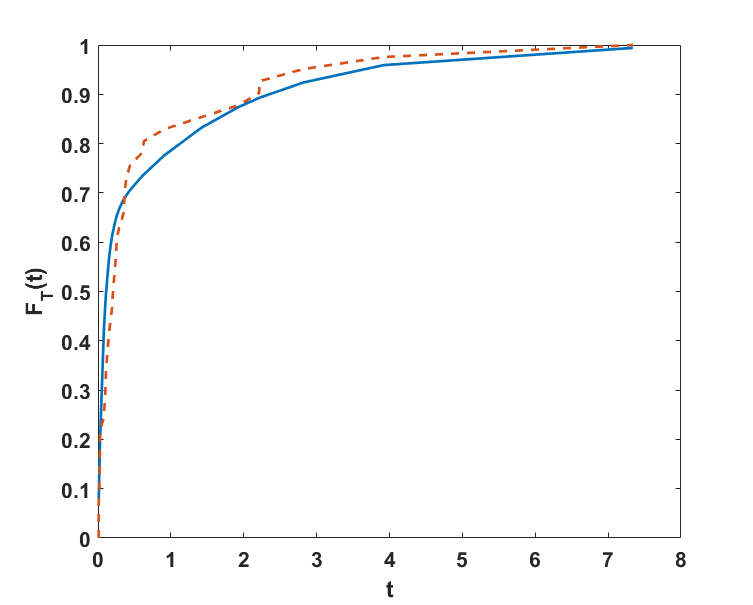}&
				\includegraphics[height=1.7in]{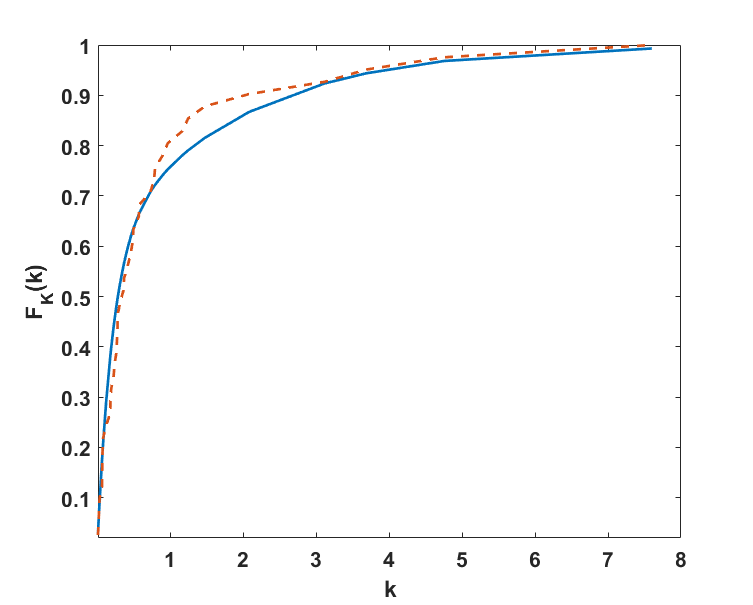}
			\end{tabular}
		\end{center}
		\caption{Estimated cdf (dashed line) under the bivariate $\mmpp_2$ versus the empirical cdf (solid line) of the inter-failures times (left panel) and distances (right panel) for train 35.}\label{fig:ft35}
	\end{figure}

	The fitting approach described in Section \ref{fittingbiv} was applied to the dataset described in Section \ref{sec:studytrains}. In particular, Table \ref{ta:trenes} and Figure \ref{fig:ft35} show the performance of the  inference method. Table \ref{ta:trenes} describes the empirical and estimated values of the expected time and covered distance between consecutive failures, as well other marginal and joint moments. The fit of the  correlation coefficient between the inter-failures times and distances is also shown. On the other hand, Figure \ref{fig:ft35} shows the fit to the empirical distribution function of the inter-failure times, $P(T<t)$, and covered distances between failures, that is, $P(K<k)$, see expression  (\ref{eq:distMap}). 
	
		\begingroup 
	\centering
	\begin{table}[H]
		\sisetup{ 
			detect-all,
			table-number-alignment = center,
			table-figures-integer = 1,
			table-figures-decimal = 3,
			explicit-sign
		} 
		\resizebox{0.9\textwidth}{!}{%
			\begin{tabular}{r|cc|cc}
				\hline
				& \multicolumn{2}{c|}{Train 35} & \multicolumn{2}{c}{Train 36} \\ 
				& Emp & Est biv $\mmpp$ & Emp & Est biv $\mmpp$ \\ 
				\hline
				\cellcolor{white} $\mu_T(1)$& \tablenum[table-format=2.2e2]{71.12} & \tablenum[table-format=2.2e2]{71.07} & \tablenum[table-format=2.2e2]{64.73}& \tablenum[table-format=2.2e2]{63.83}\\
				$\mu_T(2)$& \tablenum[table-format=2.2e2]{2.33E+04}& \tablenum[table-format=2.2e2]{2.35E+04} &  \tablenum[table-format=2.2e2]{1.11E+04} & \tablenum[table-format=2.2e2]{1.07E+04}\\
				$\mu_T(3)$& \tablenum[table-format=2.2e2]{1.25E+07} & \tablenum[table-format=2.2e2]{1.24E+07} & \tablenum[table-format=2.2e2]{2.64E+06}& \tablenum[table-format=2.2e2]{2.71E+06}\\
				$\rho_T(1)$& \tablenum[table-format=2.2e2]{0.13} & \tablenum[table-format=2.2e2]{0.13}& \tablenum[table-format=2.2e2]{0.21}& \tablenum[table-format=2.2e2]{0.19}\\
				$\mu_K(1)$& \tablenum[table-format=2.2e2]{8.78E+03} & \tablenum[table-format=2.2e2]{8.79E+03}& \tablenum[table-format=2.2e2]{8.44E+03}& \tablenum[table-format=2.2e2]{8.23E+03}\\
				$\mu_K(2)$& \tablenum[table-format=2.2e2]{2.90E+08} & \tablenum[table-format=2.2e2]{2.88E+08}& \tablenum[table-format=2.2e2]{1.83E+08}& \tablenum[table-format=2.2e2]{1.79E+08}\\
				$\mu_K(3)$& \tablenum[table-format=2.2e2]{1.58E+13} & \tablenum[table-format=2.2e2]{1.59E+13}&  \tablenum[table-format=2.2e2]{5.68E+12} & \tablenum[table-format=2.2e2]{5.81E+12} \\
				$\rho_K(1)$& \tablenum[table-format=2.2e2]{0.12} & \tablenum[table-format=2.2e2]{0.12}& \tablenum[table-format=2.2e2]{0.26}& \tablenum[table-format=2.2e2]{0.20}\\
				$\eta_{11}$& \tablenum[table-format=2.2e2]{2.54E+06} & \tablenum[table-format=2.2e2]{2.52E+06} & \tablenum[table-format=2.2e2]{1.36E+06}&
				\tablenum[table-format=2.2e2]{1.22E+06}\\
				$\eta_{21}$& \tablenum[table-format=2.2e2]{1.33E+06} & \tablenum[table-format=2.2e2]{1.30E+06} & \tablenum[table-format=2.2e2]{3.22E+08} &
				\tablenum[table-format=2.2e2]{2.95E+08} \\
				$\eta_{12}$& \tablenum[table-format=2.2e2]{1.44E+11} & \tablenum[table-format=2.2e2]{1.40E+11} & \tablenum[table-format=2.2e2]{4.18E+10} & 
				\tablenum[table-format=2.2e2]{3.94E+10} \\
				\hline 
				$Corr(T,K)$& \tablenum[table-format=2.2e2]{0.97} & \tablenum[table-format=2.2e2]{0.96}& \tablenum[table-format=2.2e2]{0.93} & \tablenum[table-format=2.2e2]{0.82}\\\hline
				\hline
		\end{tabular}}
		\vspace{0.05cm} 
		\centering 
		\caption{Empirical and estimated moments by the bivariate two-state MMPP2s for trains 35 and 36.} \label{ta:trenes}
		{\footnotesize
			\parbox{6.2in}{
				\medskip
				\begin{center}
		\end{center}}}
	\end{table}
	\endgroup
	
	\subsection{Reliability analysis}\label{reliability_data}
	Since the main interest of the analysis of the real dataset was on predicting the reliability of the system, this section is devoted to provide an assortment of measures of interest. Define $N_T(t)$ as the number of failures occurred up to time $t$, $N_K(k)$ the number of failures occurred until a distance $k$ is covered and let the joint quantity $N_{(T,K)}(t,k)$ represent the number of cumulative failures observed up to time $t$ and distance $k$. Then, the following quantities (either for train 35 and/or 36) have been predicted:
	
\begin{enumerate}
	    \item [1.] Conditional probabilities of observing a failure before a number of days $t$, assuming that a failure has occurred before $k$ kilometers are covered: $P(T<t | K<k)$.
	
	    \item [2.] Expected number of cumulative failures up to $t$ days: $E (N_T(t)).$
	    \item [3.] \textcolor{black}{Expected number of failures for different joint time and distance intervals $[t,t+dt]\times[k,k+dk]$: $E[N_{(T,K)}(t+ dt, k +dk) - N_{(T, K)}(t,k)]$, where $E(N_{(T,K)}(t,k))$ is the expected number of cumulative failures up to $t$ days and $k$ kilometers.}
	    \item [4.] Probabilities of observing $n$ failures up to time $t$, for an assortment of values of $t$ and $n$: $P(N_T(t)=n)$.
	    \item [5.] Probabilities of observing $n$ failures up to distance $k$, for an assortment of values of $k$ and $n$: $P(N_K(k)=n)$.
	    \item [6.] Joint probability of not having failures in future intervals: As it is written, this is the probability of no failures from time and km = 0 until time t and km k. This is good for a new train equal to the one you considered for the analysis. If you think of a future interval for train 35 then you should write $P(N_{(T,K)}(t+ dt, k +dk) - N_{(T, K)}(t,k)=0)$.
	\end{enumerate}
	
	Since we do not have closed formulae for the previous quantities of interest (the distribution of the counting processes associated to the bivariate $\mmpp_2$ is an open question), all predictions have been obtained through simulation. Once the estimated bivariate $\mmpp_2$ model is obtained, then traces of the same size as the real ones ($n=47$ for train 35 and $n=51$ for train 36) are simulated 1000 times. Estimations of the reliability measures are average values of the sampled data.
	
\begin{figure}[h!]
		\begin{center}
			\begin{tabular}{c c}
				\includegraphics[height=1.7in]{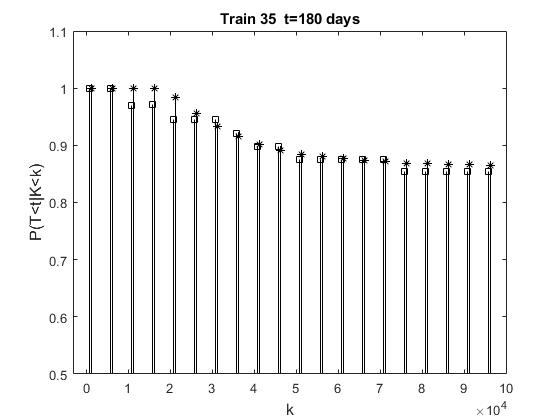}&
				\includegraphics[height=1.7in]{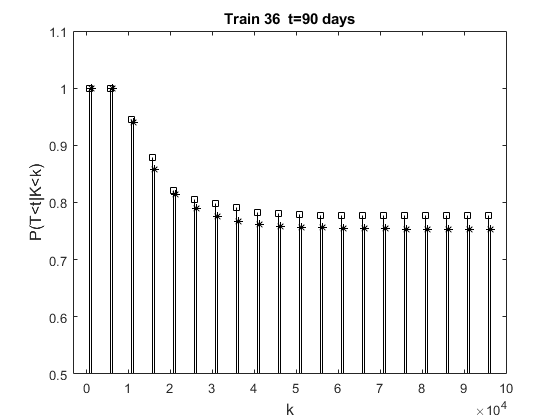}
			\end{tabular}
		\end{center}
		\caption{Comparison between the estimated (asterisk) and empirical (square) values of the conditional probabilities $p(T<t|K<k)$ for train 35 (left panel) and 36 (right panel).}\label{fig:ptk}
	\end{figure}

Figure \ref{fig:ptk} depicts the estimates of the conditional probabilities $P(T\mid K)$ in asterisk symbols. Their empirical counterparts (observed frequencies) are depicted in square symbols. The proximity between the estimated and empirical values strengthens the good fit provided by the model. The left panel of Figure \ref{fig:ptk} concerns train 35, where the probability of having a failure in less than approximately six months given that a failure was observed in less than $k\in[0,10000]$ km is estimated. The analogous probability is estimated in the right panel, for train 36 and in a period less than 3 months. The probabilities shown in Figure \ref{fig:ptk} are of interest from an engineering viewpoint since they allow to infer how likely is to observe a failure in less than a specific time period (to be fixed by the engineer), given that the distance covered until the next failure is known to be less than a given number of $km$s.

\begin{figure}[h!]
		\begin{center}
			\begin{tabular}{c c}
				\includegraphics[height=1.7in]{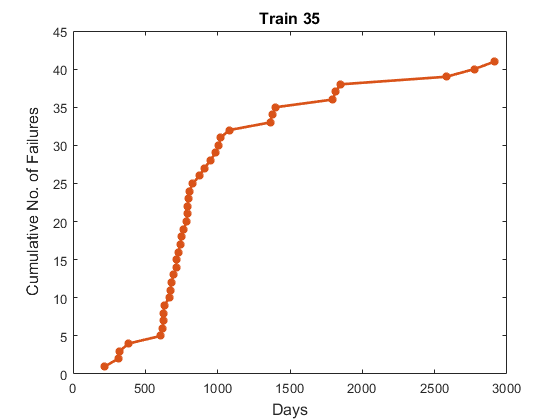}&
				\includegraphics[height=1.7in]{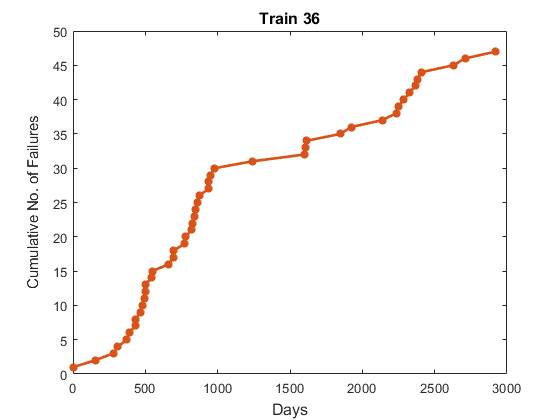}
			\end{tabular}
		\end{center}
		\caption{Estimated average number of failures under the bivariate $\mmpp_2$ for train 35 (left panel) and train 36 (right panel).}\label{fig:EntT}
	\end{figure}

Another measure of interest, from the reliability of the system viewpoint, is depicted by Figure \ref{fig:EntT}: the estimated expected number of failures in different intervals, for trains 35 and 36. From the figure, one can observe that train 35 is subject to an abrupt increased failure rate after the initial period and then heavy intervention likely occurred since the failure rate went down. Train 36 is subject to an increase in the failure rate which  decreases later (both changes occur over long periods of time). 

\begin{figure}[h!]
		\begin{center}
		\includegraphics[width=0.7\textwidth]{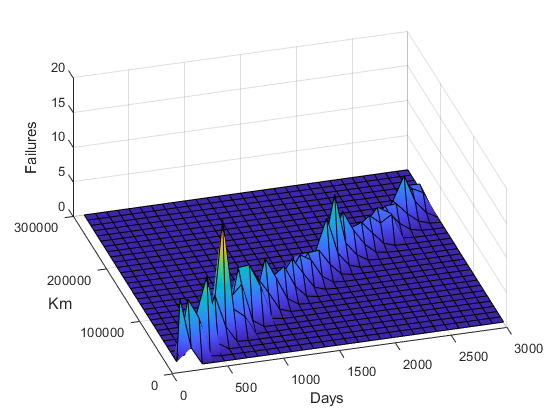}
		\end{center}
		\caption{Estimation of the expected number of failures in different intervals under the bivariate $\mmpp_2$ for train 36.}\label{fig:EstTK}
	\end{figure}

Other quantity of interest is the expected number of failures in joint intervals of time$\times$distance. Figure \ref{fig:EstTK} depicts such estimated values \textcolor{black}{of the expected number of failures in different joint time and distance intervals for train 36}. It is interesting to note that a similar behavior to that observed from the Figure was obtained in the previous studies by \cite{pievatolo2003,pievatolo2010}.

\begin{figure}[h]
		\centering
		\subfloat[ ]{{\includegraphics[width=5cm]{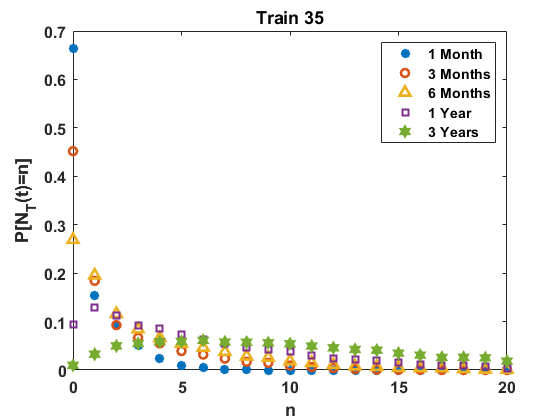}}}
		\subfloat[]{{\includegraphics[width=5cm]{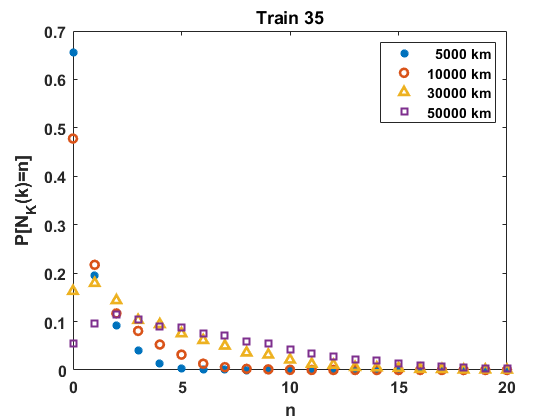} }}%
		
		\subfloat[]{{\includegraphics[width=5cm]{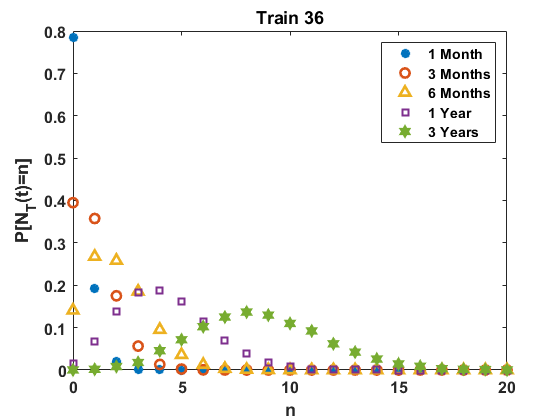} }}%
		\subfloat[]{{\includegraphics[width=5cm]{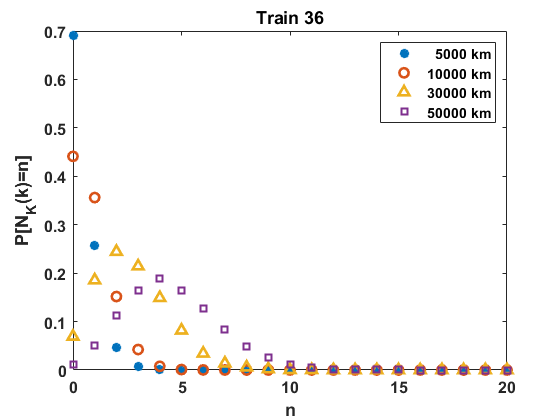} }}%
		\caption{Marginal probability functions of the number of cumulated failures for trains 35 (top panels) and 36 (bottom panels).}\label{fig:PtPk}
	\end{figure}

We also provide estimates for the marginal probability functions of the cumulated number of failures ($N_T(t)$ and $N_K(k)$). In our context, {\it marginal} is understood as referred to a single component, either time or distance. Figure \ref{fig:PtPk} shows the estimated probability functions for train 35 (top panels) and train 36 (bottom panel) for different values of cumulated failures ($n$) and an assortment of temporal periods (number of months) and distance ranges (number of kilometers).


Finally, we consider estimation for the joint probability of no failures, an information that is not available in the previous works \cite{pievatolo2003,pievatolo2010}. Tables \ref{ta:nofail35} and \ref{ta:nofail36} show the estimated probabilities of no failure for several joint intervals $[0,T]\times[0,K]$, for trains 35 and 36 respectively. It is interesting to note how for a given covered distance, the probability of zero failures decreases with time. However, for a fixed time, the considered middle distance intervals are those associated with the highest probability of no failures.

\begingroup 
	\centering
	\begin{table}[!h]
		\sisetup{ 
			detect-all,
			table-number-alignment = center,
			table-figures-integer = 1,
			table-figures-decimal = 3,
			explicit-sign
		} 
		\resizebox{\textwidth}{!}{%
			\begin{tabular}{|r|c|c|c|c|c|}
				\hline
			\backslashbox{$T$}{$K$}	& $10^5$ km &  $5\times10^5$ km & $10\times10^5$ km & $30\times10^5$ km & $50\times10^5$ km \\ 
				\hline
				\cellcolor{white} 10 Days & \tablenum[table-format=1]{1} & \tablenum[table-format=1.3]{0.659} & \tablenum[table-format=1.3]{0.480}& \tablenum[table-format=1.3]{0.170}&
				\tablenum[table-format=1.3]{0.056}\\
				\cellcolor{white} 30 Days & \tablenum[table-format=1.3]{0.665} & \tablenum[table-format=1.3]{0.607} & \tablenum[table-format=1.3]{0.474}& \tablenum[table-format=1.3]{0.170}&
				\tablenum[table-format=1.3]{0.056}\\
				\cellcolor{white} 90 Days & \tablenum[table-format=1.3]{0.452} & \tablenum[table-format=1.3]{0.452} & \tablenum[table-format=1.3]{0.438}& \tablenum[table-format=1.3]{0.167}&
				\tablenum[table-format=1.3]{0.055}\\
				\cellcolor{white} 180 Days & \tablenum[table-format=1.3]{0.270} & \tablenum[table-format=1.3]{0.270} & \tablenum[table-format=1.3]{0.269}& \tablenum[table-format=1.3]{0.164}&
				\tablenum[table-format=1.3]{0.055}\\
				\cellcolor{white} 360 Days & \tablenum[table-format=1.3]{0.103} & \tablenum[table-format=1.3]{0.103} & \tablenum[table-format=1.3]{0.102}& \tablenum[table-format=1.3]{0.102}&
				\tablenum[table-format=1.3]{0.053}\\
				\hline
		\end{tabular}}
		\vspace{0.05cm} 
		\centering 
		\caption{Estimated joint probability of no failure for different joint intervals $[0,T]\times[0,K]$ for train 35.} \label{ta:nofail35}
{\footnotesize
			\parbox{6.2in}{
				\medskip
				\begin{center}
		\end{center}}}
	\end{table}
	\endgroup

	\begingroup 
	\centering
	\begin{table}[!h]
		\sisetup{ 
			detect-all,
			table-number-alignment = center,
			table-figures-integer = 1,
			table-figures-decimal = 3,
			explicit-sign
		} 
		\resizebox{\textwidth}{!}{%
			\begin{tabular}{|r|c|c|c|c|c|}
				\hline
			\backslashbox{$T$}{$K$}	& $10^5$ km &  $5\times10^5$ km & $10\times10^5$ km & $30\times10^5$ km & $50\times10^5$ km \\ 
				\hline
				\cellcolor{white} 10 Days & \tablenum[table-format=1]{1} & \tablenum[table-format=1.3]{0.689} & \tablenum[table-format=1.3]{0.432}& \tablenum[table-format=1.3]{0.070}&
				\tablenum[table-format=1.3]{0.012}\\
				\cellcolor{white} 30 Days & \tablenum[table-format=1.3]{0.783} & \tablenum[table-format=1.3]{0.651} & \tablenum[table-format=1.3]{0.408}& \tablenum[table-format=1.3]{0.067}&
				\tablenum[table-format=1.3]{0.011}\\
				\cellcolor{white} 90 Days & \tablenum[table-format=1.3]{0.381} & \tablenum[table-format=1.3]{0.381} & \tablenum[table-format=1.3]{0.346}& \tablenum[table-format=1.3]{0.057}&
				\tablenum[table-format=1.3]{0.009}\\
				\cellcolor{white} 180 Days & \tablenum[table-format=1.3]{0.132} & \tablenum[table-format=1.3]{0.132} & \tablenum[table-format=1.3]{0.131}& \tablenum[table-format=1.3]{0.045}&
				\tablenum[table-format=1.3]{0.008}\\
				\cellcolor{white} 360 Days & \tablenum[table-format=1.3]{0.015} & \tablenum[table-format=1.3]{0.015} & \tablenum[table-format=1.3]{0.015}& \tablenum[table-format=1.3]{0.015}&
				\tablenum[table-format=1.3]{0.005}\\
				\hline
		\end{tabular}}
		\vspace{0.05cm} 
		\centering 
		\caption{Estimated joint probability of no failure for different joint intervals $[0,T]\times[0,K]$ for train 36.} \label{ta:nofail36}
{\footnotesize
			\parbox{6.2in}{
				\medskip
				\begin{center}
		\end{center}}}
	\end{table}
	\endgroup

	
	\section{Conclusions}\label{biv_conclusions}
	A bivariate extension of the two-state Markov modulated Poisson process is considered in this paper. This process allows for the modeling of non-exponential bivariate traces presenting inter- and intra-dependence, properties that make the model suitable either in reliability or other real contexts. Some properties concerning the novel model are shown, in particular the identifiability, inherited from the marginal processes, and crucial if inference is to be undertaken. 
	
	Once the process is properly described, a fitting approach is presented. The method combines a matching moment approach with an ABC algorithm. The first step helps alleviating the computational cost inherent in the ABC since the number of parameters to be estimated go from 8 to 2.
	
	The methodology is illustrated for both simulated and real datasets. In particular, an application to real failures dataset concerning a public transport company is presented. Once data are fitted by the bivariate $\mmpp_2$, a number of quantities of interest to get insight about the reliability of the system are estimated. The results show the potential of the proposed bivariate model in the reliability context as well as the correct performance of the fitting method.
	
	Prospects regarding this work concern both theoretical and applied issues. Some theoretical problems to be considered are as follows. First, we aim to derive closed  expressions for quantities of interest as the  joint probabilities in terms of  ($T$) and  ($K$) or the joint predictive distributions. Also, it is of interest to obtain probabilities of the counting process $N_T(t)$ ($N_K(k)$), number of failures  up to  $t(k)$, as in \cite{NeutsLi}. A third theoretical problem is the extension of the process proposed in this paper to its batch counterpart, where failures can occur in simultaneous way (see \cite{Yera,yera2019} for results concerning the batch counterpart of the $\mmpp_2$). Also, characterizing the bivariate $\mmpp_2$ by a set of moments is a challenging goal from the inference point of view. From an applied viewpoint, it is of interest to develop a more sophisticated version of the ABC algorithm, so that a low number of parameters is sampled, but where the existing dependence between the estimates from Step 1 and Step 2 is mitigated. It would be important also to estimate how many failures will occur in future time and km intervals, both in terms of expected number and probability of no failure in a given [time, km] interval. Work on these issues is underway.
\textcolor{black}{Finally we want to emphasize the generality of our approach.
The important formulas in Section~\ref{sec:mmpp} apply to any choice of bivariate exponential distribution, actually the formulas apply to general MPH$^\star$ distributions of arbitrary dimensions.
This generality also applies to the dependence structure that is only restricted by the general restrictions on the initial vector and the matrices of the model.
In case we have had more relevant variables in the train data set we could have included those in the modeling with only a slight additional numerical complexity.
Our model is a novel model that can be applied to reliability models with sequences of correlated multivariate observations.}

	\section*{Acknowledgments}
		The authors would like to thank the anonymous referees whose comments and suggestions improved notably the paper. This research has been financed by research projects FQM-329 and P18-FR-2369 (Junta de Andaluc\'{\i}a, Spain); PID2019-110886RB-I00,  PID2019-104901RB-I00 (Ministerio de Econom\'{\i}a, Industria y Competitividad, Spain); PID2019-104901RB-I00 (Ministerio de Ciencia e Innovación); PR2019-029 (Universidad de C\'adiz, Spain); Comunidad de Madrid through a Catedra de Excelencia. This support is gratefully acknowledged.

	\section*{Appendices}
	\appendix

	\section{Proof of Theorem \ref{Theo:BiMMPPInent}}\label{ap:Theo3}
	It is clear that if the representations  $B$ and $\bar{B}$ are equivalent, their respective marginal $MMPP_2$ will also be equivalent. Since \cite{Ryd96} proves that the $\mmpp$ is identifiable except by permutations of states, the  probabilities associated with the underlying process for the two marginals are the same (except by permutation); that is, 
	\begin{equation}
		a=\tilde{a}, \quad b=\tilde{b},
	\end{equation}
	and the failure occurrence rates of the marginal processes also coincide, (except by permutation of the vector $\boldsymbol{\lambda}$ by the vector $\boldsymbol{\omega}$), which implies that, 
	\begin{equation}\label{eq_indet_lamdas2}
		\lambda_j+\lambda_3=\tilde{\lambda}_j+\tilde{\lambda}_3, \quad \omega_j+\omega_3=\tilde{\omega}_j+\tilde{\omega}_3,  \quad \textrm{ for } j=\{1,2\}.
	\end{equation}
	On the other hand, from (\ref{eq_identifiability}) it can be proven that the bivariate exponential distribution associated with $B$ and $\tilde{B}$ are equally distributed. Note that taking $n=1$ in (\ref{eq_identifiability}), the following equality is obtained:
	\begin{equation}\label{eq:expidentifiability}
		(T_1,K_1)\stackrel{d}{=}(\tilde{T}_1,\tilde{K}_1).	
	\end{equation}
	$(T_1,K_1)$ can be rewritten as the sum  of $N$ bivariate exponential distribution, where $N-1$ is the number of times the underlying process changes state before the first failure occurs. On the other hand, since the underlying process is the same for both processes (\ref{eq:expidentifiability}) is equivalent to
	\begin{equation}\label{eq:expsum}
		(X_1+...+X_N,Y_1+...+Y_N)\stackrel{d}{=}(\tilde{X}_1+...+\tilde{X}_N,\tilde{Y}_1+...+\tilde{Y}_N),	
	\end{equation}
	and it is possible to take conditional distribution on the initial state and the number of changes state of the underlying process in (\ref{eq:expsum}) obtaining 
	\begin{equation*}
		(X_1+...+X_N,Y_1+...+Y_N|N=1,s_0=i)\stackrel{d}{=}(\tilde{X}_1+...+\tilde{X}_N,\tilde{Y}_1+...+\tilde{Y}_N|N=1,s_0=i).	
	\end{equation*}
	Hence the bivariate exponential distributions associated with $B$ and $\tilde{B}$ have to be equally distributed and consequently they have the same moments. Therefore, using (\ref{eq_indet_lamdas2}) and (\ref{eq_EXY}), the following relationship is obtained 
	\begin{equation}\label{eq_indet_lamdas3}
		\lambda_1+\lambda_2+\lambda_3=\tilde{\lambda}_1+\tilde{\lambda}_2+\tilde{\lambda}_3, \quad  \omega_1+\omega_{2}+\omega_{3}=\tilde{\omega}_1+\tilde{\omega}_2+\tilde{\omega}_3.
	\end{equation}
	Then, from (\ref{eq_indet_lamdas2}) and (\ref{eq_indet_lamdas3}), $\lambda_i=\tilde{\lambda}_i$ and $\omega_i=\tilde{\omega}_i$, for $i=\{1...3\}$, which implies the identifiability of the process in the terms defined in the paper.

			\section{Equivalent Matrix Representation}\label{ap:MatRepBiv}
	An alternative matrix representation for the bivariate $\mmpp$ is given by the initial probability vector 
	(\ref{eq_phi}):
	$$\boldsymbol{\phi}=\left(\phi_1\frac{\lambda_2}{\lambda} , \phi_1 \frac{\lambda_1}{\lambda},\phi_1 \frac{\lambda_3}{\lambda}, \phi_2\frac{\omega_2}{\omega},\phi_2\frac{\omega_1}{\omega}, \phi_2\frac{\omega_3}{\omega}\right),$$
	where $\lambda=\lambda_1 + \lambda_2 + \lambda_3$ and $\omega = \omega_1 + \omega_2 + \omega_3$, and the matrices
	$$\boldsymbol{D_0}=\left( \begin{array}{ccc|ccc}
	-\gamma_{t1} & 0 & \gamma_{t1} & 0 & 0 & 0 \\
	0 & -\gamma_{k1} & \gamma_{k1} & 0  & 0 & 0\\ 
	0 & 0 & -\lambda & \frac{\lambda}{\omega} \omega_2a & \frac{\lambda}{\omega}\omega_1 a & \frac{\lambda}{\omega}\omega_3 a
	\\&&&&&
	\\
	\hline
	&&&&&\\
	0 b & 0 & 0 & -\gamma_{t2} & 0 & \gamma_{t2}\\ 
	0& 0 & 0 & 0 & -\gamma_{k2} & \gamma_{k2}\\
	\frac{\omega}{\lambda}\lambda_2 b& \frac{\omega}{\lambda}\lambda_1 b & \frac{\omega}{\lambda}\lambda_3 b & 0 & 0 &-(\omega)
	\end{array}\right),$$
	$$\boldsymbol{D_1}=\left( \begin{array}{ccc|ccc}
	0 & 0 & 0 & 0 & 0 & 0 \\
	0 & 0 & 0 & 0 & 0 & 0\\ 
	(1-a)\lambda_2 & (1-a)\lambda_1 & (1-a)\lambda_3 & 0 & 0 & 0\\
	\hline
	0 & 0 & 0 &0 & 0 & 0\\ 
	0 & 0 & 0 & 0 & 0 & 0\\
	0 & 0 & 0 & \omega_2 (1-b) & \omega_1 (1-b) & \omega_3 (1-b)
	\end{array}\right)$$
	and
	$$\boldsymbol{R}=\left( \begin{array}{cc}
	1 & 1 \\
	0 & 1 \\ 
	0 & 1 \\
	1 & 1 \\ 
	0 & 1 \\
	1 & 1
	\end{array}\right). $$
	Note that the  parameters associated with this alternative representation are the same than for the previous one but they offer  a different arrangement in the matrix representation.
	
\section{A simulation study}\label{simu_biv}
	The aim of this section is to illustrate the behavior of the procedure described in Section \ref{fittingbiv} on the basis of two simulated datasets. Each simulated dataset consists of a sequence of 1000 pairs of failures  ($\boldsymbol{t},\boldsymbol{k}) = \{(t_1, k_1), (t_2, k_2),  . . . , (t_n, k_n)\}$ simulated from two different bivariate $\mmpp_2$s,  whose parameter sets $\{a,b,\boldsymbol{\lambda},\boldsymbol{\omega}\}$ are listed in Table \ref{ta:res_st1} in the \textit{Generator Process} columns. 
	
	The first example considers a simulated sample from a bivariate $\mmpp_2$ with low intra-dependence (both marginal processes present a first lag autocorrelation coefficient around 0.2) and very high inter-dependence (correlation between times and distances around 0.9). The second considered trace is from a bivariate $\mmpp_2$ with relatively high intra-dependence (with an autocorrelation for both marginal processes of around 0.4) and a moderate inter-dependence (correlation between times and distances around 0.7). The results obtained after running the fitting approach are shown in Table \ref{ta:res_st1}.

	\begingroup 
	\begin{table}[!h]
		\sisetup{ 
			detect-all,
			table-number-alignment = center,
			table-figures-integer = 1,
			table-figures-decimal = 3,
			explicit-sign
		} 
		\resizebox{0.9\textwidth}{!}{%
			\begin{tabular}{|c|c|c|c|c|}
				\hline
				&\multicolumn{2}{c|}{Example 1 }&\multicolumn{2}{c|}{Example 2}\\
				\hline
				&  \multicolumn{1}{c|}{$\begin{array}{c} Generator\\ Process\end{array}$}&  \multicolumn{1}{c|}{\emph{Estimation}} &    \multicolumn{1}{c|}{$\begin{array}{c} Generator\\ Process\end{array}$}&  \multicolumn{1}{c|}{\emph{Estimation}}  \\ 
				\hline
				&&&&\\
				$a$&\multicolumn{1}{c|}{0.02}&
				\multicolumn{1}{c|}{0.02}&0.008&0.008\\
				$b$&\multicolumn{1}{c|}{0.44}&
				\multicolumn{1}{c|}{0.44}&0.08&0.09\\
				&&&&\\
				$\lambda_1$&
				\multicolumn{1}{c|}{0.82}&
				\multicolumn{1}{c|}{0.68}&\multicolumn{1}{c|}{4.11}&\multicolumn{1}{c|}{4.28}\\
				$\lambda_2$&\multicolumn{1}{c|}{0.40}&
				\multicolumn{1}{c|}{0.31}&\multicolumn{1}{c|}{1.79}&\multicolumn{1}{c|}{1.62}\\
				$\lambda_3$&\multicolumn{1}{c|}{1.86}&
				\multicolumn{1}{c|}{1.84}&\multicolumn{1}{c|}{5.95}&\multicolumn{1}{c|}{5.66}\\
				&&&&\\
				$\omega_1$&
				\multicolumn{1}{c|}{\tablenum[table-format=1.2e2]{2.35e-2}}&
				\multicolumn{1}{c|}{\tablenum[table-format=1.2e2]{2.51e-2}}&\multicolumn{1}{c|}{0.12}&\multicolumn{1}{c|}{0.12}\\
				$\omega_2$&
				\multicolumn{1}{c|}{\tablenum[table-format=1.2e2]{5.27e-3}}&
				\multicolumn{1}{c|}{\tablenum[table-format=1.2e2]{7.91e-3}}&\multicolumn{1}{c|}{0.12}&\multicolumn{1}{c|}{0.12}\\
				$\omega_3$&
				\multicolumn{1}{c|}{0.24}&
				\multicolumn{1}{c|}{0.22}&\multicolumn{1}{c|}{0.33}&\multicolumn{1}{c|}{0.33}\\
				&&&&\\
				
				$\mu_T(1)$&
				\multicolumn{1}{c|}{0.58 (0.57)}&
				\multicolumn{1}{c|}{0.57}&
				\multicolumn{1}{c|}{0.29 (0.28)}&
				\multicolumn{1}{c|}{0.28}\\
				$\mu_T(2)$&\multicolumn{1}{c|}{1.99 (1.92)}&
				\multicolumn{1}{c|}{1.92}&
				\multicolumn{1}{c|}{0.88 (0.87)}&
				\multicolumn{1}{c|}{0.87}\\
				$\mu_T(3)$&\multicolumn{1}{c|}{20.16 (20.40)}&
				\multicolumn{1}{c|}{20.42}&
				\multicolumn{1}{c|}{5.67 (5.63)}&
				\multicolumn{1}{c|}{5.65}\\
				$\rho_T(1)$&\multicolumn{1}{c|}{0.22 (0.21)}&
				\multicolumn{1}{c|}{0.22}&
				\multicolumn{1}{c|}{0.41 (0.40)}&
				\multicolumn{1}{c|}{0.40}\\
				&&&&\\
				$\mu_K(1)$&
				\multicolumn{1}{c|}{0.66 (0.66)}&
				\multicolumn{1}{c|}{0.65}&
				\multicolumn{1}{c|}{0.31 (0.32)}&
				\multicolumn{1}{c|}{0.32}\\
				$\mu_K(2)$&
				\multicolumn{1}{c|}{2.38 (2.30)}&
				\multicolumn{1}{c|}{2.31}&
				\multicolumn{1}{c|}{0.89 (0.89)}&
				\multicolumn{1}{c|}{0.90}\\
				$\mu_K(3)$&
				\multicolumn{1}{c|}{25.37 (25.78)}&
				\multicolumn{1}{c|}{25.75}&
				\multicolumn{1}{c|}{5.72 (5.79)}&
				\multicolumn{1}{c|}{5.77}\\
				$\rho_K(1)$&
				\multicolumn{1}{c|}{0.21 (0.22)}&
				\multicolumn{1}{c|}{0.21}&
				\multicolumn{1}{c|}{0.39 (0.40)}&
				\multicolumn{1}{c|}{0.39}\\
				&&&&\\
				$\eta_{11}$&
				\multicolumn{1}{c|}{2.02 (1.99)}&
				\multicolumn{1}{c|}{1.93}&
				\multicolumn{1}{c|}{0.69 (0.70)}&
				\multicolumn{1}{c|}{0.69}\\
				$\eta_{21}$&\multicolumn{1}{c|}{20.10 (19.26)}&
				\multicolumn{1}{c|}{19.98}&
				\multicolumn{1}{c|}{3.84 (3.90)}&
				\multicolumn{1}{c|}{3.85}\\
				$\eta_{12}$&\multicolumn{1}{c|}{21.27 (21.87)}&
				\multicolumn{1}{c|}{21.16}&
				\multicolumn{1}{c|}{3.85 (3.81)}&
				\multicolumn{1}{c|}{3.88}\\
				&&&&\\
				$Cor(T,K)$&\multicolumn{1}{c|}{0.91 (0.90)}&
				\multicolumn{1}{c|}{0.90}&
				\multicolumn{1}{c|}{0.76 (0.75)}&
				\multicolumn{1}{c|}{0.76}\\
				&&&&\\
				$\begin{array}{c} running\\ time\end{array}$
				&\multicolumn{1}{c|}{-}&\multicolumn{1}{c|}{219.79}&-&\multicolumn{1}{c|}{213.89}\\			\hline \hline
		\end{tabular}}
		\vspace{0.05cm} 
		\centering 
		\caption{Comparison between the theoretical, empirical (within parentheses) and estimated values for obtained with the algorithm of Table 2 for two examples.} \label{ta:res_st1}
		{\footnotesize
			\parbox{6.2in}{
				\medskip
				\begin{center}
		\end{center}}}
	\end{table}
	\endgroup

	The first eight rows in Table \ref{ta:res_st1} show the parameters of the process (the real ones under the column \textit{Generator Process} and the estimated ones under the column \textit{Estimated}). The ninth to sixteenth rows depict the marginal empirical moments that characterize marginal processes (theoretical and empirical in parenthesis in the \textit{Generator Process} column and estimated ones in the \textit{Estimation }column). The analogous can be found from row 17th to row 19th, but regarding the joint moments. The penultimate row shows the correlation between \textcolor{black}{inter-failure} times and distances, and finally, the last row shows the running time (measured in seconds) in an Intel Xeon Six-cores 3.6 GHz with 12 threads  processor with 128Gb of memory ram (for a prototype code written in MATLAB${}^\copyright$).

	Some comments arise from the results presented in Table \ref{ta:res_st1}. On one hand, it should be pointed out the good performance of the method for estimating the moments of the bivariate process (both marginal and joint) and in particular, the correlation between the inter-failure times and distances. Parameters $a$ and $b$ are also well fitted, a fact that also happens for $\lambda_3$ and $\omega_3$ since the ABC approach is specifically designed in terms of them. However, it has been observed that estimations for the rest of parameters ($\lambda_1,\lambda_2, \omega_1$ and $\omega_2$) may be less good for some cases. Consider for example, the first simulated trace. From Step 1 in the fitting approach, the values of $\gamma_{t1}=\lambda_1+\lambda_3$ and $\gamma_{k1}=\lambda_2+\lambda_3$  (equal to 2.53 and 2.26) are estimated as 2.53 and 2.25. Since the generated values of $\lambda_3$ in Step 2 (ABC) are upper-bounded by $\min\{\hat{\gamma}_{t1},\hat{\gamma}_{k1}\}$, then the results under the ABC approach will be better or worse depending on the estimates for $\gamma_{t1}$ and $\gamma_{k1}$. In the previous example, the final estimation for $\lambda_2$ turns out slightly better than that of $\lambda_1$, as expected. A similar fact occurs in the case of the parameter $\omega_3$. 
	
	An additional comment regards the computational time of the proposed fitting approach. For the considered examples, the total running time is around 6 minutes, where the most of computational cost is due to the ABC algorithm. The matching moment approach defining Step 1 is not expensive from a computational point of view (see \cite{yera2019}). However, it is known from the literature that the ABC algorithm is time consuming, in particular for high values of $I_2$ (number of iterations or number of times where traces are simulated), see \cite{minter2019}. In our case, such value, $I_2$, is set as 10000 which provides a good compromise between performance of the inference approach and computing time.

\begin{figure}[h!]
		\begin{center}
			\begin{tabular}{c c}
				\includegraphics[height=1.7in]{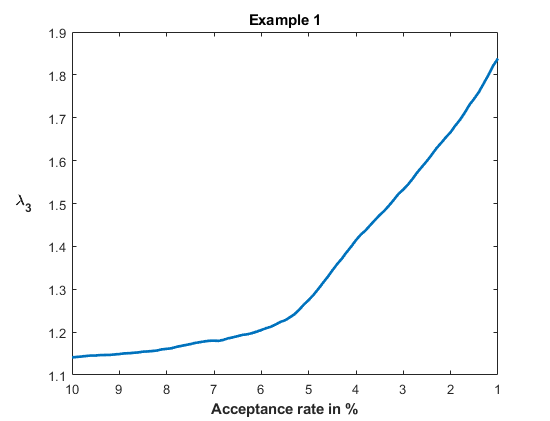}&
				\includegraphics[height=1.7in]{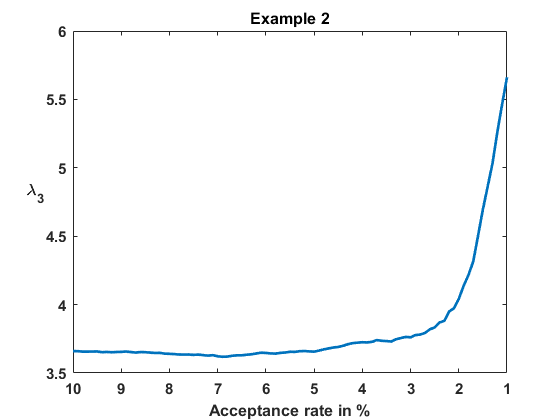}
			\end{tabular}
		\end{center}
		\caption{Evolution of the estimated value $\lambda_3$ for different acceptance percentages in the ABC algorithm (Step 2 in the fitting approach). The left (right) panel refers to the first (second) simulated example.}\label{fig:Lambdas}
	\end{figure}

	In order to explore in more depth the results under the ABC algorithm, consider Figure \ref{fig:Lambdas} and Table \ref{ta:res_st1}. Figure \ref{fig:Lambdas}  shows the evolution of the estimation of parameter $\lambda_3$ as the acceptance percentage considered in the ABC algorithm varies.  Note that in left panel (Example 1) the estimated value with 10\% acceptance is $1.14$, which is very close to  the mean value of the prior distribution of $\hat{\lambda}_3$ ($\min(\hat{\gamma}_{t1},\hat{\gamma}_{k1})/2$). As the acceptance rate decreases, $\hat{\lambda}_3$ approaches to $\lambda_3$, as expected. A similar behavior can be seen in the right panel (Example 2). Table \ref{ta:res_porcent} shows the whole set of parameters $(\boldsymbol{\lambda}, \boldsymbol{\omega})$ for different acceptance rates. Again, it can be seen that as the acceptance percentage decreases, the estimate becomes more accurate. 
	
	\begingroup 
	\begin{table}[!h]
		\sisetup{ 
			detect-all,
			table-number-alignment = center,
			table-figures-integer = 1,
			table-figures-decimal = 3,
			explicit-sign
		} 
		\resizebox{0.8\textwidth}{!}{%
			\begin{tabular}{|c|c|c|c|c|c|c|}
				\hline
				&\multicolumn{6}{c|}{Example 1}\\
				\hline
				&  \multicolumn{1}{c|}{$\lambda_1$}&\multicolumn{1}{c|}{$\lambda_2$} &    \multicolumn{1}{c|}{$\lambda_3$}&  \multicolumn{1}{c|}{$\omega_1$}&\multicolumn{1}{c|}{$\omega_2$}&\multicolumn{1}{c|}{$\omega_3$}  \\ 
				\hline
				&&&&&&\\
				\multicolumn{1}{|c|}{$\begin{array}{c} Generating \\ process\end{array}$}&
				0.82 & 0.40 & 1.86 & 0.02 & 0.005 & 0.24 \\
				&&&&&&\\
				$1\%$& 0.69 & 0.32 & 1.84 & 0.03 & 0.008 & 0.22  \\
				$5\%$ & 1.25 & 0.88 & 1.27 & 0.02 & 0.007 & 0.22 \\
				$10\%$ & 1.39 & 1.01 & 1.14 & 0.03 & 0.01 & 0.21  \\
				
				&&&&&&\\
				
				\hline
				&\multicolumn{6}{c|}{Example 2}\\
				\hline
				&  \multicolumn{1}{c|}{$\lambda_1$}&\multicolumn{1}{c|}{$\lambda_2$} &    \multicolumn{1}{c|}{$\lambda_3$}&  \multicolumn{1}{c|}{$\omega_1$}&\multicolumn{1}{c|}{$\omega_2$}&\multicolumn{1}{c|}{$\omega_3$}  \\ 
				\hline
				&&&&&&\\
				\multicolumn{1}{|c|}{$\begin{array}{c} Generating \\ process\end{array}$}&
				4.11 & 1.79 & 5.95 & 0.12 & 0.12 & 0.33 \\
				&&&&&&\\
				$1\%$& 4.28 & 1.62 & 5.66 & 0.12 & 0.12 & 0.33 \\
				$5\%$ & 6.29 & 3.63 & 3.66 & 0.12 & 0.12 & 0.33 \\
				$10\%$ & 6.28 & 3.63 & 3.66 & 0.12 & 0.12 & 0.33 \\
				&&&&&&\\
				\hline \hline
		\end{tabular}}
		\vspace{0.05cm} 
		\centering 
		\caption{Estimates for $\boldsymbol{\lambda}$ and $\boldsymbol{\omega}$ when varying the acceptance percentage in the ABC algorithm (Step 2 in the fitting approach)} \label{ta:res_porcent}
		{\footnotesize
			\parbox{6.2in}{
				\medskip
				\begin{center}
		\end{center}}}
	\end{table}
	\endgroup

	\bibliographystyle{myapalike}
	\bibliography{ref_paper1}

\end{document}